\documentclass{article}%
\usepackage{amsmath}
\usepackage{amsfonts}
\usepackage{amssymb}
\usepackage{graphicx}
\usepackage{subcaption}
\usepackage{multirow}
\usepackage{color}
\usepackage{longtable}
\usepackage[dvipsnames]{xcolor}
\usepackage[bookmarks=true,colorlinks=true]{hyperref}
\hypersetup{linkcolor=blue}
\hypersetup{citecolor=blue}
\hypersetup{urlcolor=blue}
\newcommand{\bpartial}{\mathop{\partial\kern -4pt\raisebox{.8pt}{$|$}}}
\newcommand{\bra}{\mathopen{[\kern-1.6pt[}}
\newcommand{\ket}{\mathclose{]\kern-1.5pt]}}
\newcommand{\bbra}{\mathopen{[\kern-2.2pt[\kern-2.3pt[}}
\newcommand{\bket}{\mathclose{]\kern-2.1pt]\kern-2.3pt]}}

\makeindex
\begin{document}
	\title {\large{ \bf 
			Spatially homogeneous black hole solutions in $z=4$ Ho\v{r}ava-Lifshitz gravity in $(4+1)$ dimensions with Nil geometry and $H^2\times  R$  horizons}}
	
	\vspace{3mm}
		\author {  \small{ \bf  F. Naderi}\hspace{-1mm}{ \footnote{e-mail:
			 f.naderi@azaruniv.ac.ir (corresponding author)}},
		 { \small	} \small{ \bf  A. Rezaei-Aghdam}\hspace{-1mm}{
		\footnote{ e-mail:	rezaei-a@azaruniv.ac.ir}},
	{ \small	} \small{ \bf Z. Mahvelati-Shamsabadi}\hspace{-1mm}{		\footnote{ e-mail:	 z.mahvelati@azaruniv.ac.ir}} 
	\\
		 		{\small{\em
			Department of Physics, Faculty of Basic Sciences, Azarbaijan Shahid Madani University}}\\
	{\small{\em   53714-161, Tabriz, Iran  }}}

\maketitle

\begin{abstract}
In this paper, we present two new families of spatially homogeneous black hole solution for $z=4$ Ho\v{r}ava-Lifshitz Gravity equations in  $(4+1)$ dimensions with  general coupling constant $\lambda$ and the especial case $\lambda=1$, {considering $\beta=-1/3$}. The three-dimensional horizons are considered to have Bianchi types $II$ and $III$ symmetries, and hence the horizons are modeled on two types of Thurston $3$-geometries, namely the Nil geometry and $H^2\times R$. Being foliated by compact 3-manifolds,  the horizons are neither spherical, hyperbolic, nor toroidal, and therefore are not of the previously studied topological black hole solutions in Ho\v{r}ava-Lifshitz gravity. Using the Hamiltonian formalism, we establish the conventional thermodynamics of the solutions defining the mass and entropy of the black hole solutions for several classes  of solutions. It turned out that for both horizon geometries the area term in the entropy receives two non-logarithmic negative corrections proportional to Ho\v{r}ava-Lifshitz parameters. Also, we show that choosing some proper set of parameters the solutions can exhibit locally stable or unstable behavior.

\end{abstract}

\section{Introduction}
\subsection{General considerations}
The non-relativistic power
counting renormalizable theory of   Ho\v{r}ava-Lifshitz gravity was proposed by Ho\v{r}ava at the Lifshitz point
aimed at resolving the problems concerning the ultraviolet behavior of Einstein gravity \cite{Hoava2009,Hoava2,Horava3}. Ho\v{r}ava-Lifshitz gravity explicitly breaks the Lorentz invariance and restores
Einstein’s general relativity at low-energy limits \cite{Afshordi2009}. This modified theory of gravity, which preserves spatial general covariance and time reparametrization invariance,  can be regarded as a good candidate for presenting a quantum field theory of gravity \cite{Wang2017}. 

Ho\v{r}ava-Lifshitz gravity has received growing interest and a large number of studies have explored the implications of this proposal in detail. For instance, the  cosmological solutions of Ho\v{r}ava-Lifshitz gravity have been studied in \cite{Kiritsis_2009,Brandenberger_2009,Nojiri2010,Christodoulakis2012,Giani2017}, cosmological perturbation and the related properties have been discussed in \cite{Mukohyama12009,Piao2009,Gaoo2019,Wangg2010,Wang32010,Wang62009,Cai992010,Cognola2016},  and some other properties of Ho\v{r}ava-Lifshitz gravity have been investigated in \cite{Saridakis2010,Bogdanoss2010,Leonlo2009,Chagoyaya2019}. 
Particularly,  much attention has been paid to 
black hole solutions and their thermodynamics behavior
in the  framework of Ho\v{r}ava-Lifshitz gravity   \cite{Park2009,GHODSI2011a,Ghodsi2010,L2009,Cai22009,Caia2009,Cai2009,Blas2011,Barausse2011,Eling2016}. In this context, for instance, the quantum gravity effects by using Ho\v{r}ava-Lifshitz black hole have been investigated in \cite{Pourhassan2018}, phase transition and  the quasinormal modes of a massive scalar field in the background of a rotating Ho\v{r}ava AdS black hole was analyzed in \cite{Bcar2020,Chen_2010}, and  properties of the black hole solutions were researched in \cite{Cai2009,Myung2010,Chenn2009,Pengg22010,Chench2009,Lee2010}.  

The  Ho\v{r}ava-Lifshitz solutions are usually classified by the anisotropy degree between space and time, indicated by the so-called $z$ parameter.
	Particularly, the $z = 3$ case has attracted much attention  for which the theory is a non-relativistic renormalizable gravity model at short distance, providing a candidate
	quantum field theory of gravity in the UV \cite{Cai22009}. Many works have been done in  $z = 3$ Ho\v{r}ava-Lifshitz gravity, such as black hole solutions \cite{Park2009,L2009,Caia2009},
	Hawking radiation \cite{Pengg22010}, thermodynamical properties \cite{Myung2010,Wang}, perturbation \cite{Wang32010}, and observational effects \cite{Abdujabbarov}. 
	Soon after intruding the  original Ho\v{r}ava-Lifshitz gravity,  the z = 4 Ho\v{r}ava-Lifshitz gravity was proposed in \cite{Cai22009}, studied on $(4+1)$ and $(3+1)$ dimensions for instance in \cite{Horava3,Park2009,Cai22009,Chench2009,Liu2013}.
	Despite its importance,  the $z= 4$ case has been studied much less extensively than $z=3$. One of the main motivations to consider this case is its importance in $(3+1)$ dimensions, where from the viewpoint of spectral dimension, the $z=4$ is favorable because of its consistency
	with the results of lattice quantum gravity numerical
	simulations  \cite{Horava3,Ambj}.  On the other hand, in $(4 + 1)$ dimensions, power counting super renormalizability  in UV region requires $z = 4$ \cite{Hoava2}.  Further
	discussions supporting $z=4$ case have also been presented \cite{Hoava2,Cai22009,Poshteh}.

In this work, we are going to  consider  $ z=4 $   Ho\v{r}ava-Lifshitz gravity in $(4+1)$ dimensions,  searching  for new topological  black hole solutions. The topological black hole solutions in Ho\v{r}ava-Lifshitz gravity were first found in \cite{Caia2009}. 
So far, the horizon geometries of topological black hole solutions in  Ho\v{r}ava-Lifshitz theory in $(4+1)$ and $(3+1)$ dimensions have been considered to be spherical, hyperbolic, or flat, indicated by the constant scalar curvature of the horizon, namely $k=1,-1$, and $0$. 
However, in $(4+1)$ dimensions the situation can be more extensive, where the event horizon of a stationary black hole can be 
orientable compact  $3$-dimensional Riemann manifolds, which are required to be
endowed with a metric.
Based on Thurston geometrization conjecture \cite{thurston1983three},  proved later by {Perelman} \cite{perelman2003ricci}, the geometry of such $3$-manifolds
is locally isometric to one
of the eight Thurston type geometries, including three isotropic constant  scalar curvature
cases of spherical $S^3$, Hyperbolic $H^3$, and  Euclidean $E^3$, product constant curvature types  $S^2 \times R$, $H^2 \times
R$, and twisted product types of    $\widetilde{{SL_2R}}$, Solve geometry, and Nil geometry \cite{thurston1983three}.  Except for $S^3$ and $E^3$, the other Thurston type geometries are negatively curved spaces. All of these model geometries admit homogeneous metrics and show a close  correspondence with the  
Bianchi types and Kantowski-Sachs homogeneous models
\cite{Closed,closedBianchi}.
The homogeneous spacetimes, possessing a symmetry called the spatial homogeneity  \cite{Cosmictopology}, have been widely used in finding cosmological solution  in the context of  Einstein gravity    \cite{Ellis1969}, string theory  \cite{batakis2,PhysRevD.57.5108,NADERI2017,Naderi2,naderi2020classical}, and Ho\v{r}ava-Lifshitz gravity theory  \cite{Christodoulakis2012,Giani2017}.

Particularly interesting families of $(4+1)$ dimensional black hole 
solutions of some gravity theories have been presented in the framework of Bianchi type spacetimes,  where the horizons are modeled by some of the Thurston  $3$-geometries  \cite{Iizuka2012,Cadeau,Liu2012,PhysRevD.91.084054,PhysRevD.97.024020,HERVIK20081253,Naderi2019,Nil1,Nil2}.
  These types of black holes are especially of interest in the context of AdS/CFT and holography approaches, where the generators of the translational symmetry are generalized to  Bianchi symmetries to avoid some complications for theories in $(3+1)$ dimensions \cite{Iizuka2012,Nil1}. 
	So far, no black hole solution with Thurston horizon geometries has been obtained for the Ho\v{r}ava-Lifshitz gravity theory.
	Thus, considering Ho\v{r}ava-Lifshitz  gravity as a candidate quantum gravity theory and the importance of investigating 
	AdS/CFT  correspondence in the framework of this theory \cite{Cai,Cai066003},  it is interesting to
	find black hole solutions with special Thurston type horizon geometries for $(4+1)$ dimensional  Ho\v{r}ava-Lifshitz gravity, for which  the power counting super renormalizbality requires $z=4$.

In this paper, we are interested in  spatially homogeneous black hole solutions for $z=4$ Ho\v{r}ava-Lifshitz gravity on $(4+1)$ dimensional spacetimes, where the three-dimensional horizons are particularly assumed to be homogeneous spaces corresponding to  Bianchi types $II$ and $III$ with closed geometries of  Nil and $H^2\times R$, respectively. These negatively curved homogeneous geometries are non-trivial in the sense that they are not  constant scalar curvature 
type geometries that have been extensively studied in previous topological black hole solutions in Ho\v{r}ava-Lifshitz gravity.

The paper is organized as follows: In section \ref{sec2}, a review on Ho\v{r}ava-Lifshitz gravity and its action {for $z=4$ case in $(4+1)$ dimensions}  is presented. 
In section \ref{sec3}, we obtain topological black hole solution for the equations of motion of $z=4$ Ho\v{r}ava-Lifshitz gravity on $(4+1)$ dimensional spacetimes,  whose horizons corresponding to the Bianchi types $II$ and $III$ homogeneous spaces, have Nil geometry and $H^2\times R$ geometry, respectively. Then, the thermodynamic behavior of the solutions is investigated in section \ref{sec4}.  Finally, some concluding remarks are presented in section \ref{conclussion}.

\section{Brief review on Ho\v{r}ava-Lifshitz gravity}\label{sec2}
In this section, we present some introductory remarks on $z = 4$ Ho\v{r}ava-Lifshitz gravity  \cite{Hoava2009,Hoava2,Cai22009}. 
On $(D+1)$ dimensional spacetime, the $ADM$ metric decomposition  can be considered as follows  
\begin{equation}\label{metric1}
	\begin{split}
		dS^{2}=-N^{2}dt^{2}+g_{ij}(dx^{i}-N^{i}dt)(dx^{j}&-N^{j}dt),i,j=1,...,D,
	\end{split}
\end{equation}
where $N$, $N^i$, and $g_{ij}$ are, respectively,   the lapse function,  shift function, and spatial metric. 
The Ho\v{r}ava gravity models exhibit an anisotropic time and space scaling invariance, given by
\begin{equation}
	x^i\rightarrow l x^i, ~ ~ ~ ~t\rightarrow l^{z}t ,
\end{equation}
where the dynamical critical exponent $ z $ indicates the degree of anisotropy between space and time. 
Under this transformation  $g_{ij}$ and $N$ are invariant, but $N^i$ is scaled as $N^i\rightarrow l^{1-z} N^i$. 
In  the 
inverse spatial length units, the dimensions of time  and space in the Ho\v{r}ava-Lifshitz gravity are  $[t] = -z,$ $ [x] = -1,$ $ [c] = z-1$, at the fixed point with Lifshitz index $z$. The Ho\v{r}ava-Lifshitz gravity   in $z=1$ case   yields  the familiar general relativity in the IR limit.
In the UV region,  renormalizablity of Ho\v{r}ava-Lifshitz theory requires 
different values of  $z$, where  the theory becomes power-counting
renormalizable  with $z_{UV} = D$, and super-renormalizable with $z_{UV} > D$.

The simplest kinetic for Ho\v{r}ava-Lifshitz gravity is given by \cite{Hoava2009,Hoava2,Cai22009}
\begin{equation}
	\label{60}
	S_{K}=\frac{2}{\kappa^2}\int dt d^Dx\sqrt{g}N(K_{ij}K^{ij}-\lambda K^2), 
\end{equation}
where $ g $ is  determinant of the $D$-dimensional metric $ g_{ij} $,  and
\begin{equation}\label{Kij}
	K_{ij}=\frac{1}{2N}(\dot{g_{ij}}-\nabla _i N_j -\nabla _j N_i),
\end{equation}
is the extrinsic curvature associated with the spatial metric,  $  K=g^{ij}K_{ij} $ is its trace, 
$ \kappa $ is a coupling constant with the scaling dimension  at the fixed point $ [\kappa]=\frac{z-D}{2} $ that is dimensionless in $D= z=4 $ case,  and $ \lambda $ is a dimensionless parameter. Particularly, the   $ \lambda=1 $ restores the kinetic term of Einstein's theory. 

The potential term, which  satisfies the so-called "detailed balance condition", is given by  \cite{Hoava2009,Hoava2}
\begin{equation}
	\label{606}
	S_{V}=\frac{\kappa^2}{8}\int dt d^Dx\sqrt{g}N E^{ij} {\cal{G}}_{ijkl} E^{kl}, 
\end{equation}
where
\begin{equation}{\cal{G}}_{ijkl}=\frac{1}{2}(g_{ik}g_{jl} + g_{il}g_{jk}) -\tilde{\lambda}  g_{ik}g_{jl},\quad \tilde{\lambda}=\frac{\lambda}{D \lambda -1},
\end{equation}
is the inverse of  DeWitt supermetric, defined by 
$ {\cal{G}}^{ijkl} =\frac{1}{2}(g^{ik}g^{jl} + g^{il}g^{jk}) - {\lambda} g^{ik}g^{jl} $
where ${\cal{G}}_{ijmn}{\cal{G}}^{mnkl}=\frac{1}{2}(\delta_i^k\delta_j^l+\delta_i^l\delta_j^k)$. Also, $E^{ij}$ coming from the $D$-dimensional relativistic action \cite{Hoava2009,Hoava2}
\begin{equation}\label{E}
	E^{ij} =\frac{1}{\sqrt{g}}\frac{\delta W_{D}[g_{kl}]}{\delta g_{ij}},
\end{equation}
is the detailed balance condition, which establishes the connection between $D$-dimensional system described by the
action $W_D$ to a $(D +1)$ dimensional system described by the action $S_K -S_V$. 
A theory with spatial isotropy would require $W _{D}$ to be the action of relativistic theory in Euclidean signature.

\subsection{Action for $z=4$ Ho\v{r}ava-Lifshitz gravity in $(4+1)$ dimensions }
In this paper, our focus will be on $z=4$ Ho\v{r}ava-Lifshitz gravity in $(4+1)$ dimensional  spacetimes, where the theory  is power-counting renormalizable. In this case,  $4$-dimensional relativistic Lagrangian is given in the following general form \cite{Hoava2,Cai22009}
\begin{equation}\label{w4}
	\begin{split}
		W_4=&\frac{1}{ k_w^2} \int  d^4x\sqrt{g}\left(R-2\Lambda _W\right)+\frac{1}{M}\int  d^4x\sqrt{g}\left(R_{ij}R^{ij}+\beta R^2\right),
	\end{split}
\end{equation}
in which $ k_w $, $ \Lambda _W $, $ M$, and $ \beta$ are coupling constants and $ R_{ij} $ and $ R $ are the Ricci tensor and Ricci scalar, respectively.
Noting that the Gauss-Bonnet combination is a topological invariant in four dimensions, the second term  in \eqref{w4} includes the most general form  of curvature square contribution. 
Now, according to \eqref{E}, $E^{ij}$ is given by
\begin{equation}\label{e4}
	E^{ij}=-\frac{1}{ k_w^2} (G^{ij}+\Lambda _W g^{ij})-\frac{1}{M}L^{ij},
\end{equation}
where
\begin{equation}
	G^{ij}= R^{ij}-\frac{1}{2}g^{ij}R,
\end{equation}
\begin{equation}\label{Lij}
	\begin{split}
		L^{ij}=&(1+2\beta)(g^{ij}\nabla ^{2}-\nabla ^i\nabla ^j)R+\nabla ^2G^{ij}
		+2\beta R(R^{ij}-\frac{1}{4}g^{ij}R)+2(R^{imjn}-\frac{1}{4}g^{ij}R^{mn})R_{mn}.
	\end{split}
\end{equation}
Then, combining the kinetic  and  potential terms, the $z=4$  Ho\v{r}ava-Lifshitz gravity in $(4+1)$ dimensions is given by  the following  Lagrangian \cite{Cai22009,Li014} 
\begin{equation}
	\label{440}
	{\cal{L}}={\cal{L}}_0+{\cal{L}}_1,
\end{equation}
\begin{equation}
	{\cal{L}}_0=\sqrt{g}N\left(\frac{2}{\kappa^2}\left(K^{ij}K_{ij}-\lambda K^2\right)+\frac{\kappa^2(\Lambda _W R-2\Lambda _W^2)}{4 k_w^4 (1-4\lambda)}\right),
\end{equation}
\begin{eqnarray}
	\begin{split}
		{\cal{L}}_1=-\sqrt{g}N&\frac{\kappa^2}{8}\bigg(\frac{1}{ k_w^4} G_{ij}G^{ij}+\frac{2}{M k_w^2}G^{ij}L_{ij}+\frac{2}{M k_w^2}\Lambda _W L+\frac{1}{M^2}L^{ij}L_{ij}\\
		&-\tilde{\lambda}\left(\frac{L^2}{M^2}-\frac{2 L}{M k_w^2}\left(R-4\Lambda _W\right)+\frac{1}{k_w^4}R^2\right)\bigg),
	\end{split}
\end{eqnarray}
in which 
\begin{equation}
	L=2(1+3\beta)\nabla ^2 R. 
\end{equation}
In order to restore general relativity in the IR region, the  relations between the effective couplings and the speed of light $c$,  Newton coupling $G$, and the effective cosmological constant $\Lambda$ are emerged as
\begin{equation}\label{d}
	c=\frac{ \kappa^2}{k_w^2\sqrt{8}}\sqrt{\frac{\Lambda _w}{1-4 \lambda}}, \quad
	G_N=\frac{ \kappa^2 c}{32\pi},\quad
	\Lambda =\Lambda _W.
\end{equation}
Then, in  IR region, the $\lambda=1$ case gives rise to 
general relativity,  provided that the  $\Lambda _w$ takes negative value to have a  well-defined $c$. Then, with   negative $\Lambda_W$, reality of physical parameters in \eqref{d} needs $\lambda>\frac{1}{4}$.

\section{Topological $(4+1)$ dimensional black hole solutions for $z = 4$ Ho\v{r}ava-Lifshitz gravity }\label{sec3}

We are looking for black hole solutions of $z=4$ Ho\v{r}ava-Lifshitz gravity equations of motion on $(4+1)$ dimensional spacetime, where the $r$ and $t$ constant hypersurfaces {will be assumed to be } given by homogeneous spaces corresponding to  Bianchi types $II$ and $III$. There is a correspondence between geometries of Bianchi type $II$ and $III$ symmetric spaces and the Thurston type Nil and $H^2\times R$ geometries  \cite{thurston1983three}, respectively, where  $H^2$ denotes two-dimensional hyperbolic space.
The former space is a twisted product manifold while the latter one is a product of constant curvature manifolds.

Setting $N^i=0$ in \eqref{metric1},  we start with the following metric ansatz 
\begin{equation}\label{metricc}
	ds^2=-(N(r))^2 f(r) dt^2+\frac{dr^2}{f(r)}+g_{\alpha\beta}(r)\sigma^{\alpha}\sigma^{\beta},
\end{equation}
where $\alpha=1, 2, 3$, and the $\sigma^{\alpha}$ are left invariant  $1$-form basis of Bianchi types, given by \cite{batakis2,Ryan} 
\begin{equation}\label{ii}
	{ II:}\quad	\sigma^1=dx^2-x^1dx^3,\quad \sigma^2=dx^3,\quad \sigma^3=dx^1,
\end{equation}
\begin{equation}\label{iii}
	{III:}\quad	\sigma^1=dx^1,\quad \sigma^2=dx^2,\quad \sigma^3={\rm e}^{x^1}dx^3.
\end{equation}
As long as the metric coefficients are independent of $x^i$, the metric will be automatically invariant
under the Bianchi type isometrics \cite{Iizuka2012,Ryan}. The horizons for these topological black holes are negatively  curved spaces that can not be described by
Einstein spaces metric, i.e.  $R_{\alpha\beta}=k g_{\alpha\beta}$.

It is difficult to obtain solutions for the general value of $\beta$ in \eqref{440}. We will restrict our attention to  special value $\beta=-\frac{1}{3}$.  Following the method of \cite{L2009,Caia2009}, we will obtain the solutions by substituting the
metric ansatz into the Ho\v{r}ava-Lifshitz action with the Lagrangian \eqref{440}.

\subsection{Solution in Bianchi type II}\label{IIsolutions}

In this Bianchi type, noting \eqref{metricc} and \eqref{ii}, the metric ansatz can be considered as follows
\begin{equation}\label{metricII}
	\begin{aligned}
		ds^2=-(N(r))^2 f(r) dt^2&+\frac{dr^2}{f(r)}+a r^n\left(dx^2-x^1dx^3\right)^2+b r^m\left(\left(dx^1\right)^2+\left(dx^3\right)^2\right),
	\end{aligned}
\end{equation}
where $n$, $m$, $a$ and $b$ are constants. The $a$ and $b$ constants 
introduce eventual additional scales. The components of $R_{ij}$, $R$, $K_{ij}$, and $L_{ij}$ for this case are presented in Appendix  \ref{app1}, considering $\beta=-\frac{1}{3}$.
It is quite difficult to find the exact solution for general values of $n$ and $m$. Interestingly, setting $m=n$  leaves only the two derivative terms in the Ho\v{r}ava-Lifshitz action\footnote{It is worth mentioning that  for black hole solutions with Nil geometry horizon the requirement of metric of type \eqref{metricII} to admit an additional isometry  corresponding to Lifshitz scale 	invariance, constraints the $n$ and $m$ constants to $n=2m$ \cite{Iizuka2012,PhysRevD.91.084054}. In this work, we will consider only the $3$-isometries of Bianchi types, letting $n=m$. However, if the Lifshitz scaling invariance on the horizon is the case of interest with $m=n$, it can be admitted by setting $a=\sqrt{r_H^{-n}}$.}
\begin{eqnarray}\label{actionII}
	\begin{split}
		I=&\int dtdx^4\frac { \kappa^2aN}{{b}^{3} k_w^2} \bigg( 
		\frac {{a}^{2}{r}^{-\frac{n}{2}}}{72 {b}^{4}{M}^{2}} \big( 3 { 
			{{b}^{4}M{ n} \left( -{ f'} r+2 {f} \right) }{{r}^{-2}}
		}-44 {a}^{2} k_w^2{r}^{-2 { n}}+33 {b}^{2}M{a}{r}
		^{-{ n}} \big) \\
		&+\frac {1}{32 k_w^2 \left( 4
			\lambda-1 \right) }\big(\left( 3 {
			n}-4 \lambda-2 \right){a}   {b}^{2}{ n} {r}^{\frac{ n}{2}-2}{f}+3 \left( 3 {{ n}}^{2}-4 
		(\lambda-1)-6 { n}\right) {b}^{4}{{ n}}^{2}{r}^{
			\frac{3}{2} { n}-4}{{f}}^{2}\\
		&+3 \left(  \left( 8 \Lambda {f}-{{f'}}^{2
		} \left( \lambda-1 \right)  \right) { n}-8 \Lambda {f}
		\right) {b}^{4}{ n} {r}^{ \frac{3}{2} { n}-2}+3 {b}^{4}{{ n}}^{2} \left( 3 { n}-4+4 \lambda \right) 
		{r}^{ \frac{3}{2} { n}-3}{f'} {f}\\
		&+{ n} {r}^{-1+
			\frac{n}{2}} {b}^{2}{f'}\left( {a} \left( 2 \lambda+1 \right) +12 \Lambda {
			b}^{4}{r}^{{ n}} \right) + \left( 3-11 \lambda \right) {a
		}^{4}{r}^{-\frac{n}{2}}+16 {r}^{ \frac{3}{2} { n}}{b}^{4}{\Lambda}^{2}+4 
		{r}^{\frac{n}{2}}{b}^{2}\Lambda {a}\big) \bigg),
	\end{split}
\end{eqnarray}
where prim, here and hereafter, stands for derivative with respect to $r$. Variation of this action with respect to $f(r)$ gives the follwoing equation of motion
\begin{eqnarray}\label{eq1II}
	\begin{split}
		4 &{\ln(N)'} \big( -3a {b}^{2}M \left( 2 \lambda+1 \right) {r}^{2-n}+4  \left( 4 
		\lambda -1\right) k_w^{2}{a}^{2}{r}^{-2 n+2}+9 {b}^{4}M ( 
		-n \left( 3 n-4+4 \lambda \right) f\\
		&+2 n{  f'}  \left( \lambda-1
		\right) r-2 \Lambda {r}^{2} ) 
		\big) +18r^{-1}  {fn{b}^{4}M \left( 3 {n}^{2}+6  \left( 1-2 \lambda
			\right) n-8+8 \lambda \right) }-8 {a}^{2} \left( 4 \lambda-1
		\right)  \left( n-2 \right)k_w^{2}{r}^{-2 n+1}\\
		&-6 {b}^{2}M
		\left(  \left( 2 \lambda-5 \right) n+4 \lambda+2 \right) a{r}^{-n+1
		}+36 {b}^{4}M ( n{  f'}  \left( 3 n-4 \right)  \left( 
		\lambda-1 \right) +2  ( n \left( \lambda-1 \right) { f''}+
		\Lambda\left( n-2 \right)  ) r ) 
		=0.	\end{split}
\end{eqnarray} 
Also, the equation of motion of $N(r)$ function can be read easily from the action \eqref{actionII}.

In order to guarantee the existence of  black
hole solutions which are not necessarily extremal, we impose the boundary conditions at the event horizon with 
$f(r_H)=0$, $f'(r_H)\neq 0$, and  finite lapse function $N(r_H)$, where the subscript $H$, here and in what follows, denotes quantities evaluated at the horizon. 
Then, the equations of motions yield the following conditions 
on the horizon
\begin{eqnarray}\label{eq1hII}
	\begin{split}
		\bigg[&		4 {\ln(N)'} \big( -3a {b}^{2}M \left( 2 \lambda+1 \right) {r}^{2-n}+4  \left( 4 
		\lambda -1\right) k_w^{2}{a}^{2}{r}^{-2 n+2}+9 {b}^{4}M \left( 
		2 n{  f'}  \left( \lambda-1
		\right) r-2 \Lambda {r}^{2} \right) 
		\big)\\
		&-8 {a}^{2} \left( 4 \lambda-1
		\right)  \left( n-2 \right)k_w^{2}{r}^{-2 n+1}-6 {b}^{2}M
		\left(  \left( 2 \lambda-5 \right) n+4 \lambda+2 \right) a{r}^{-n+1
		}\\
		&+36 {b}^{4}M ( n{  f'}  \left( 3 n-4 \right)  \left( 
		\lambda-1 \right) +2  ( n \left( \lambda-1 \right) { f''}+
		\Lambda\left( n-2 \right)  ) r )  
		\bigg]_{r=r_H}=0,
	\end{split}
\end{eqnarray}
\begin{eqnarray}\label{eq2hII}
	\begin{split}
		\bigg[ &{
			\frac {{a}^{2}{r}^{-\frac{n}{2}}}{72 {b}^{4}{M}^{2}} \left( -3 { 
				{{b}^{4}M{ n}{ f'} }{{r}^{-1}}
			}-44 {a}^{2} k_w^2{r}^{-2 { n}}+33 {b}^{2}M{a}{r}
			^{-{ n}} \right) }-3    {{f'}}^{2
		} \left( \lambda-1 \right)  { n}
		{b}^{4}{ n} {r}^{ \frac{3}{2} { n}-2}+{f'} { n} {r}^{-1+
			\frac{n}{2}} ( {a} \left( 2 \lambda+1 \right) \\
		&+12 \Lambda {
			b}^{4}{r}^{{ n}} ) {b}^{2}+ \left( -11 \lambda+3 \right) {a
		}^{4}{r}^{-\frac{n}{2}}+16 {r}^{ \frac{3}{2} { n}}{b}^{4}{\Lambda}^{2}+4 
		{r}^{\frac{n}{2}}{b}^{2}\Lambda {a}\big)
		\bigg]_{r=r_H}=0.
	\end{split}
\end{eqnarray}
It is worth mentioning that the negative Ricci scalar of the horizon in this Bianchi type is $R^{(3)}=-\frac{a}{2b^2r_H^{n}}$. Accordingly,  we define the following parameter for further uses
\begin{equation}\label{alfaII}
	\alpha\equiv\frac{a}{2b^2}.
\end{equation}

Now, solving the equations of motion, we find the following three classes of solutions:

$\bullet$  For  special case $\lambda=1$ we obtain the following  solutions 
\begin{equation}\label{lamda1N}
	N(r)= N_0r^{\frac{n}{2}-1},
\end{equation}
\begin{eqnarray}\label{lamda1fII}
	\begin{split}
		f(r)&=-  \frac{4}{3}  {\frac {{r}^{2}
				\Lambda}{{{ n}}^{2}}}-  \frac{2}{3}  {\frac {{r}^{2-n}{\alpha}}{{{ n}}^{2
		}}}+\frac{16}{9}{\frac { {{\alpha}}^{2} k_w^2 {r}^{2(1-n)}}{ M{{ n}}^{
					2}}}
		\\
		&-\frac{1}{9}{\frac {{r}^{2- n}}{Mb^8{{ n}}^{2}}}\bigg( 9 {n} {b}^{8}{M}^{2} \left( 9 {C_1} {{n}}^{3}{b}^{8
		}+16 {a}^{4}\ln  \left( r \right)  \right) +256  \left( 3 {b}^{4}M-
		2 {a}^{2} k_w^2{r}^{-{n}} \right) {a}^{6} k_w^2{
			r}^{-{n}}
		\bigg)^{\frac{1}{2}},
	\end{split}
\end{eqnarray}
in which the $N_0$ and $C_1$ are integrating constants. 
This set of solutions  satisfies the boundary conditions \eqref{eq1hII} and \eqref{eq2hII}, using $f(r_H)=0$.

When the conditions \eqref{d} hold, general relativity  in the IR region can be recovered in the $\lambda=1$ case, at $M\rightarrow \infty$ limit, noting \eqref{w4} and \eqref{e4}. Black hole solutions in the presence of a negative cosmological constant with Nil geometry horizon have been obtained for general relativity in \cite{Iizuka2012,PhysRevD.91.084054,Nil1}, however with different values for the $m$ and $n$ constants in the metric \eqref{metricII}, as a consequence of applying generalized Lifshitz scaling invariance on the metric. Here, the obtained solution for $f(r)$ function with considering only the Bianchi symmetry and setting $m=n$ in metric \eqref{metricII}, recasts the following form when $M\rightarrow \infty$ 
	\begin{eqnarray}\label{dd}
		\begin{split}
			f(r)&=-  \frac{4}{3}  {\frac {{r}^{2}
					\Lambda}{{{ n}}^{2}}}-  \frac{2}{3}  {\frac {{r}^{2-n}{\alpha}}{{{ n}}^{2
			}}}
			-\frac{1}{3}{\frac {{r}^{2- n}}{b^4}} \left( 9 {C_1} {{n}}^{3}{b}^{8
			}+16 {a}^{4}\ln  \left( r \right)  \right) 
			^{\frac{1}{2}},
		\end{split}
	\end{eqnarray}
	which compared to the solutions of \cite{Iizuka2012,PhysRevD.91.084054}  contains an extra logarithmic term, even with $n=2$. However, it is worth mentioning that \eqref{dd} may remind the Nil geometry solutions with intermediate scaling obtained in \cite{Nil1}, which contains logarithmic function at the boundary.

$\bullet$ When $\lambda$ is allowed to have any value, in the special case of $n=2$  we obtain
\begin{equation}\label{n2N}
	N(r)= N_0,
\end{equation}
\begin{equation}\label{n2ii}
	f(r)=-\frac{\alpha}{6}+C_2r^2+\frac{C_1}{r^2},
\end{equation}
where $C_1$ and $C_2$ are integrating constants. $C_2$ is actually the cosmological constant redefined up to a factor. These two functions satisfy the boundary conditions on the horizon, given by \eqref{eq1hII} and \eqref{eq2hII}.

$\bullet$ Also,  another class of solution  can be obtained for general values of $n$ and $\lambda$ as follows
\begin{equation}\label{N2II}
	N(r)= N_0,
\end{equation}
\begin{eqnarray}\label{f2II}
	\begin{split}
		f(r)=-  \frac{4}{3}  {\frac {{r}^{2}
				\Lambda}{{{ n}}^{2}}}-  \frac{2}{3}  {\frac {{r}^{2-n}{\alpha}}{{{ n}}^{2
		}}}+&\frac{16}{9}{\frac { {{\alpha}}^{2} k_w^2 {r}^{2(1-n)}}{ M{{ n}}^{
					2}}}- C_1 r^{s_1}+C_2 r^{s_2},
	\end{split}
\end{eqnarray}
in which 
\begin{equation}\label{ss}
	s_1=-\frac{3 }{4}(n-2)-{\sqrt{\mu}},\quad s_2=-\frac{3 }{4}(n-2)+{\sqrt{\mu}},
\end{equation}
and 
\begin{equation}\label{mu}
	\mu={\frac { \left( 3 n+2 \right) ^{2}\lambda-21 {n}^{2}+12 n-4}{16 
			(\lambda-1)}}.
\end{equation} 
The solutions \eqref{N2II} and \eqref{f2II} are consistent with 
the boundary conditions on the horizon. Reminding the $\lambda>\frac{1}{4}$ condition required by reality  of the speed of light in \eqref{d}, we will also exclude the  values of $\lambda$ in the range  of  $1<\lambda<{\frac {21 {n}^{2}-12 n+4}{ \left(3 n+2 \right) ^{2}}}$, for which $\mu$ is negative. Here, similar to the solutions presented in \cite{Cai2009a}, there are two branches in \eqref{f2II}. It is easy to find that $s_1$ is negative for any values of $n$ and $\lambda$, but $s_2$ can have different signs. 
Practically, with $n>2$, for $\lambda>{\frac {21 {n}^{2}-12 n+4}{ \left(3 n+2 \right) ^{2}}}$ the power of $r$ in $C_1$-dependent term is negative and in the range of $(\frac{3}{2}(2-n),-\frac{3}{2}n+1)$, while for $C_2$-dependent term the power of $r$ is in the range of $(\frac{3}{2}(2-n),2)$, which can be either positive or negative, but still less than $2$.  Also, with $n>2$, for $\frac{1}{4}<\lambda<1$ the power of $C_1$-dependent term is again negative, while for $C_2$-dependent term  the power is larger than $2$. On the other hand, for $n<2$, $s_1$ is again negative for any values of $\lambda$ but $s_2$, being positive, is less then $2$ as $\frac{1}{4}<\lambda<1$ and larger than $2$ in $\lambda>{\frac {21 {n}^{2}-12 n+4}{ \left(3 n+2 \right) ^{2}}}$ range. Dominance of $\Lambda r^2$ term in \eqref{f2II} at large distance  suggests that the solution can have  asymptotic behavior of AdS spacetime
. But, in the cases that the $C_2$-dependent term, having the power of $r$ larger than $2$,  is dominant at the large distances, the solutions may not have clear physical meaning {\cite{Cai2009a}}.
Considering this point, to investigate the thermodynamic behavior of this family of solutions we consider the following two cases:

(i) When $n$ and $\lambda$ are primarily independent and arbitrary parameters, from the asymptotic behavior point of view, similar to the solutions presented in \cite{Cai2009a} that only the negative branch solutions were selected, we will focus only on the  $C_1$-dependent term which has negative power of $r$ for any values of $\lambda$ and $n$.\footnote{However, as we will see in the following,  a well-defined mass in the Hamiltonian approach needs   $\mu=0$, which practically leaves no difference between $C_1$ and $C_2$-dependent terms in \eqref{f2II}.}

(ii) Another special case appears if,  keeping the $\lambda$ parameter general,  the following relation between $\lambda$ and $n$ holds
\begin{equation}\label{lanadII}
	\lambda=\frac{3}{4}{\frac {{ n} \left( { n}-2 \right) }{3 { n}-2}}+1.
\end{equation}
It leads to $s_2=0$ for  $n\geq2$, where $f(r)$ function recasts the following form\footnote{Similarly, for $n<2$ we have $s_1=0$, leading to the last two terms in $f(r)$ in the form of $-{ C_1}+{ C_2}{r}^{\frac{3}{2}n-3}$.}
\begin{eqnarray}\label{sol1II}
	\begin{split}
		f(r)=-  \frac{4}{3}  {\frac {{r}^{2}
				\Lambda}{{{ n}}^{2}}}-  \frac{2}{3}  {\frac {{r}^{2-n}{\alpha}}{{{ n}}^{2
		}}}+&{\frac {16  {r}^{2(1-n)}{{\alpha}}^{2} k_w^2}{9  M{{ n}}^{
					2}}}
			+{ C_2}-{r}^{-\frac{3}{2}n+3}{ C_1},
	\end{split}
\end{eqnarray}
where the $\Lambda$ term is dominant at large $r$ because the exponents of the other terms are negative. 
With $n=2$, which according to \eqref{lanadII} is accompanied with $\lambda=1$, the  $f(r)$ function turns into that of $(4+1)$ dimensional  black hole solution in $z=4$ Ho\v{r}ava-Lifshitz gravity with spherical, flat and hyperbolic horizons for $\lambda=1$ case,  presented in \cite{Liu2013}, where the thermodynamic behavior of solutions has been also discussed. Hence, we will restrict our attention  in this case on $n>2$.

It is worth adding a  remark on the asymptotic isometries of the considered classes of solutions  with Nil geometry horizon for $\lambda=1$ and general $\lambda$. In general, the obtained metrics at $r\rightarrow \infty$ contain a generalized dilatation generator whose action on the coordinates  is given as following
	\begin{eqnarray}
		\begin{split}
			t\rightarrow \rho t,\, 	r\rightarrow \rho^{-1} r,\, 	x^1\rightarrow \rho^{\frac{n}{2}} x^1,\, 	x^2\rightarrow \rho^{n} x^2,\, 	x^3\rightarrow \rho^{\frac{n}{2}} x^3,\,\,\,\,
		\end{split}
	\end{eqnarray} 
	with constant $\rho$, using a scaling in the constant $a$ of metric. 
	Note that for $n=2$, there is only one anisotropic direction $x^2$, similar to the Nil horizon black hole solution for general relativity obtained in \cite{PhysRevD.91.084054},
	while for general values of $n$, or equivalently the general values of $\lambda$, the anisotropy appears in all $x^i$ directions.

\subsection{Solutions in Bianchi type III }\label{solIII}
In this Bianchi type, noting \eqref{metricc} and \eqref{iii},  we can have  the following metric ansatz 
\begin{equation}\label{metricIII}
	\begin{split}
		ds^2=-&(N(r))^2 f(r) dt^2+\frac{dr^2}{f(r)}+a r^n \left((dx^1)^2+{\rm e}^{2 x^1} (dx^3)^2 \right)+b r^m (dx^2)^2,
	\end{split}
\end{equation}
where $n$, $m$, $a$ and $b$ are constants.  The components of $R_{ij}$, $R$, $K_{ij}$, and $L_{ij}$ for this case are presented in Appendix  \ref{app1}.
Particularly, setting $m=n$ and substituting  the metric  into the action gives
\begin{eqnarray}\label{action}
	\begin{split}
		I=\int dtdx^4 &\frac {\sqrt {b}N \kappa^2N}{{{
					k_w}}^{2}{M}^{2}{a}^{3}} \bigg[ \frac {{r}^{-\frac{n}{2}}}{72} ( 6
		{ {{a}^{2}fMn}{{r}^{-2}}}-3  { {{a}^{2}Mn{ f'}}{r^{-1}}}+12  
		aM{r}^{-n}-4 k_w^2{r}^{-2  n} )\\
		& -\frac {{r}^
			{\frac{n}{2}}}{32 k_w^2 \left(4  \lambda -1 \right) }\bigg( 4  
		n ( 4  \lambda-3  n+2 ) {a}^{3}{M}^{2}{{r}^{-2}}f-4
		{ {n{a}^{3}{ }  {M}^{2} \left( 2  \lambda+1 \right) }{r^{-1}}f'}\\
		&+
		3  {M}^{2}{f}^{2}{a}^{4}{n}^{2} {r}^{n-4} \left( -3  {n}^{2}+4  \lambda+6  n-4
		\right)-3  {n}^{2}{a}^{4}{ f'}  {M}^{2}{r}^{n-3}f \left( 3  n-4+4
		\lambda \right)\\
		&+3  n{a}^{4}{M}
		^{2}{r}^{-2+n} \left(  \left( \lambda-1
		\right) n{{ f'}}^{2}-8  \Lambda  f \left( n-1 \right)  \right) \\
		&-4  {a}^{2}{M}^{2} \big(  \left( -4  \lambda+2 \right) 
		{r}^{-n}+a\Lambda   \left( 4+a \left( 3  { f'}  {r}^{n-1}n+4  {r}^{
			n}\Lambda \right)  \right)  \big)  \bigg)  \bigg].
	\end{split}
\end{eqnarray}
Variation of this action with respect to $f(r)$ gives the equation of motion as follows
\begin{eqnarray}\label{eq1}
	\begin{split}
		\frac{1}{2}M & n{r}^{2  n}{a}^{2} \left( \lambda-1
		\right) \left(2 {f''}  {r}^{2}+{f'}   \left( 3  n-4 \right) r+4  
		f   \right)
			-\ln(N)' \big( \frac{1}{2} \left(4 \lambda+{3}n-4 \right) n{a}^{2}Mf{r}
		^{2n+1}\\
		&-M{  f'}{a}^{2}n \left( \lambda-1 \right) {r}^{2n+2}+\frac{2}{3}Ma \left( 2\lambda+1 \right) {r}^{3+n}+2\Lambda M{r}^{2 n+3}{
			a}^{2}-\frac{2}{9} \left( 4 \lambda-1 \right) {{  k_w}}^{2}{r}^{3}
		\big)
		\\
		&-\frac{1}{3} aM \left(  \left( 2\lambda-{5} \right) n+4 \lambda+2
		\right) {r}^{2+n}-\frac{1}{9}  \left( n-2 \right)  \left( 4 \lambda-1
		\right) {{  k_w}}^{2}{r}^{2}+{a}^{2}M\Lambda  \left( n-2 \right) 
		{r}^{2 n+2}\\
		&+\frac{3}{4} n \left( {n}^{2}+ \left(2 -4 \lambda\right) n
		\right) {a}^{2}fM{r}^{2n}=0,
	\end{split}
\end{eqnarray}
where the equation of motion of $N(r)$ can be easily read of the action. 
Also, on the horizon, with the  conditions $f(r_H)=0$ and $f'(r_H)\neq 0$ and finite lapse function $N(r_H)$, we should have
\begin{eqnarray}\label{eq1h}
	\begin{split}
		&\bigg[\frac{1}{2}M  n{r}^{2 n}{a}^{2} \left( \lambda-1
		\right) \left(2 {f''} {r}^{2}+{f'}  \left( 3 n-4 \right) r \right)
			-\ln(\tilde{N})' \big(-M{  f'}{a}^{2}n \left( \lambda-1 \right) {r}^{2n+2}+\frac{2}{3}Ma \left( 2\lambda+1 \right) {r}^{3+n}\\
		&+2\Lambda M{r}^{2 n+3}{
			a}^{2}-\frac{2}{9} \left( 4 \lambda -1\right) {{  k_w}}^{2}{r}^{3}
		\big)-\frac{1}{9}  \left( n-2 \right)  \left( 4 \lambda-1
		\right) {{  k_w}}^{2}{r}^{2}+{a}^{2}M\Lambda  \left( n-2 \right) 
		{r}^{2 n+2}
		\\
		&	-\frac{1}{3} aM \left(  \left( 2\lambda-{5} \right) n+4 \lambda+2
		\right) {r}^{2+n}\bigg]_{r=r_H}=0,
	\end{split}
\end{eqnarray}
\begin{eqnarray}\label{eq2h}
	\begin{split}
		&\bigg[\frac{1}{8}{  f'}{a}^{2} Mn \big( -{3} {r}^{2 n}Mn{a}^{2}  {  f'} 
		\left( \lambda-1 \right)  +12{r}^{2 n+1}{a}^{2}M\Lambda+4aM \left(2 \lambda+1
		\right) {r}^{1+n}-\frac{2}{3} {{  k_w}}^{2}r \left(4 \lambda-1
		\right)  \big)\\
		& +\frac{2}{9} {r}^{2}{{  k_w}}^{2}
		\left( 4 \lambda -1\right)  \left( 3 {r}^{-n}aM-{r}^{-2 n}{{
				k_w}}^{2} \right)+{a}^{2} M^2{r}^{2}\left( 2 {r}^{2 n}{a}^{2}{\Lambda}^{2}+2 {r}^{n}a
		\Lambda-2 \lambda+1 \right)\bigg]_{r=r_H}=0.
	\end{split}
\end{eqnarray}
The horizon geometry of the black hole solutions with Bianchi type $III$ with metric \eqref{metricIII} is equivalent to   $H^2\times R$, and the negative Ricci scalar of the horizon is $R^{(3)}=-\frac{2}{ar_H^n}$. For further uses, we define the parameter $\alpha$ in this Bianchi type by
\begin{eqnarray}
	\alpha=\frac{2}{a }. 
\end{eqnarray}
Now, solving the equations of motion
we obtain the following classes of solutions:

$\bullet$ For the special case $\lambda=1$, we obtain
\begin{equation}\label{n1}
	N(r)= N_0r^{\frac{n}{2}-1},
\end{equation}
\begin{eqnarray}\label{lamda1f}
	\begin{split}
		f(r)=&-\frac{4}{3}{\frac {{r}^{2}\Lambda}{{n}
				^{2}}}-\frac{2}{3}{\frac {{r}^{2-n}{   \alpha}}{{n}^{2}}}+\frac{1}{9}{\frac {{r}^{2(1-n)}{
					{  k_w}}^{2}{{   \alpha}}^{2}}{M{n}^{2}}}\\
		&-\frac{1}{9}\frac{r^{-n+2}}{n^2 M} \bigg(9{M}^{2}n \left( 9{ C_1}{n}^{3}+4\ln 
		\left( r \right) {\alpha}^{2} \right) +2{\alpha}^{3}{{
				k_w}}^{2}{r}^{-n} \left( 12M - k_w^2{r}^{-n}
		\alpha\right) 
		\bigg)^{\frac{1}{2}},
	\end{split}
\end{eqnarray}
which are consistent with the boundary conditions \eqref{eq1h} and \eqref{eq2h}, without any extra condition on the constants.
Similar to the solutions with Nil geometry horizon \eqref{lamda1fII}, even at $M\rightarrow \infty$, the obtained $f(r)$ function for metric \eqref{metricIII} with $m=n$, contains  logarithmic function that has  not appeared in the other solutions with $H^2\times R$ geometry horizon, where apart from  the $3$-isometries of Bianchi type $III$, the metric was required to be invariant Lifshitz generalized transformations imposing $n=0$ \cite{Cadeau}.

$\bullet$ When  $\lambda$ is allowed to have general values, with $n=2$ we find the solutions
\begin{equation}\label{n2iii}
	N(r)= N_0,
\end{equation}
\begin{equation}\label{n2iii2}
	f(r)=-\frac{\alpha}{6}+C_2r^2-\frac{C_1}{r^2}.
\end{equation}

$\bullet$ 
Also, for the general value of $\lambda$ and $n$ we obtain the solutions
\begin{equation}\label{N2}
	N(r)= N_0,
\end{equation}
\begin{eqnarray}\label{f2}
	\begin{split}
		f(r)=-\frac{4}{3}{\frac {{r}^{2}\Lambda}{{n}
				^{2}}}-\frac{2}{3}{\frac {{r}^{2-n}{   \alpha}}{{n}^{2}}}&+\frac{1}{9}{\frac {{r}^{2(1-n)}{
					{  k_w}}^{2}{{   \alpha}}^{2}}{M{n}^{2}}}- C_1 r^{s_1}+C_2 r^{s_2},
	\end{split}
\end{eqnarray}
where  the $s_1$ and $s_2$ constants are again given by \eqref{ss}.
Here, similar to what we had in the Bianchi type $II$ solutions given by \eqref{N2II} and \eqref{f2II}, we will highlight two cases. First, when the $n$ and $\lambda$ are independent, we will focus on the $C_1$-dependent term. 
In the second case, imposing the special relation \eqref{lanadII}  between $\lambda$ and $n$, we obtain the $f(r)$ function in the following form 
\begin{eqnarray}\label{fIII}
	\begin{split}
		f(r)=-\frac{4}{3}{\frac {{r}^{2}\Lambda}{{n}
				^{2}}}-\frac{2}{3}{\frac {{r}^{2-n}{   \alpha}}{{n}^{2}}}&+\frac{1}{9}{\frac {{r}^{2(1-n)}{
					{  k_w}}^{2}{{   \alpha}}^{2}}{M{n}^{2}}}	+{ C_2}-{r}^{-\frac{3}{2}n+3}{ C_1},
	\end{split}
\end{eqnarray}
in which no inconsistency arises in presence of $C_2$-term.

The obtained solutions in all three classes of $\lambda=1$, $n=2$  and general $\lambda$ for  Bianchi types $II$ and $III$ are seemed to resemble each other closely. The thermodynamic behavior of the solutions will be studied in the following, where we can  compare the physical behaviors.  
It is worth mentioning that, the solutions in these two Bianchi type classes are not of the constant curvature type with $R_{\alpha\beta}=kg_{\alpha\beta}$, and the Ricci scalar of the horizon is a function of the radius of the horizon. However, the $\alpha$ parameters, defined by $\alpha=-R^{(3)}r_H^{n}$, appeared in the solutions somehow similar to  the $k$ parameter of the topological black hole solutions with constant  curvature horizons \cite{L2009,Cai22009,Caia2009,Cai2009,Blas2011}.

All group of solutions obtained for $H^2\times R$ horizon geometry with general $\lambda$ and $\lambda=1$   are asymptotically invariant under the following
	Lifshitz generalized transformations 
	\begin{eqnarray}
		\begin{split}
			t\rightarrow \rho t,\, 	r\rightarrow \rho^{-1} r,\, 	x^1\rightarrow  x^1,\, 	x^2\rightarrow \rho^{\frac{n}{2}} x^2,\, 	x^3\rightarrow  x^3,\,\,\,\,
		\end{split}
	\end{eqnarray} 
	with constant $\rho$, if one uses the scaling of $a$.	Accordingly, for $n=2$ the solutions are asymptotically isotropic,
	while for general values of $n$, or equivalently general values of $\lambda$, the anisotropy appears in  $x^2$ direction.

\section{Thermodynamic properties of the black hole solutions}\label{sec4}
In this section we are going to  establish the thermodynamic of the obtained solutions,  using the canonical Hamilton
formulation, where noting the metric \eqref{metricc}, the Euclidean continuation of the action is given by \cite{Cai2009} 
\begin{eqnarray}\label{}
	\begin{split}
		I_E &=	\int dt dx^4 \left(\pi^{ij}\dot{g}_{ij}-\sqrt{f}N{\cal{H}}-N^i{\cal{H}}_i\right)+B,
	\end{split}
\end{eqnarray}
in which  $B$ denotes the boundary term. In our cases  $N^i=0$ and we have
\begin{eqnarray}\label{Eaction}
	\begin{split}
		I_E &=-\beta\Omega	\int_{r_+}^{\infty} dr {N}{\cal{H}}+B,
	\end{split}
\end{eqnarray}
where $\beta$ is the period of Euclidean time,  $\Omega=\int\sigma^1\times\sigma^2\times\sigma^3$
denotes the volume of $3$-dimensional closed space, and $r_+$ is the radius of the black hole outer horizon defined
by the largest root of $f(r)=0$. 
For static black holes, the constraint ${\cal{H}}=0$ is required to be satisfied and then the Euclidean action reduces to the boundary term $B$.
Namely, for the on-shell solutions we
have 
\begin{equation}\label{Ie}
	I_E=B=B\mid_{\infty} -‌‌B\mid_{r_+}.
\end{equation}
In fact, supplementing the action with boundary term ensures obtaining a well-defined variational principle on these non-asymptotically flat space-times.

Regularity of Euclidean black hole solution requires the time period $\beta$ to follow the following relation
\cite{Cai22009,Cai2009}
\begin{equation}\label{T1}
	\beta (N(r)f'(r))\mid _{r_+}=4 \pi,
\end{equation}
which yields the temperature of the black hole by
\begin{equation}\label{T2}
	T=\frac{1}{\beta}.
\end{equation}
Also, the relation between Euclidean action and  free energy $F_e$
\begin{equation}\label{free}
	I_E=\beta F_e=\beta m-S,
\end{equation}
can be used to obtain the  mass $m$ and entropy $S$ of the  black hole solutions.

\subsection{Thermodynamics of Bianchi type $II$ black hole solutions with Nil geometry horizon}

There is a correspondence between the geometry of Bianchi type $II$ spaces and Thurston's Nil geometry and Heisenberg group,
whose
isotropy groups are $SO(2)$ and $ e$, respectively \cite{closedBianchi}. 	
We have found the black hole solutions in this Bianchi type in section \ref{IIsolutions}, represented in terms of the Ho\v{r}ava-Lifshitz  constants $\kappa, k_w, M$,   $\lambda$, and the horizon curvature constant related parameter    $\alpha=\frac{a}{2b^2}$. The area of the horizon for this Bianchi type solutions is 
\begin{eqnarray}\label{Ahii}
	A_H=\sqrt{2\alpha} b^2r_+^\frac{3n}{2}\Omega.
\end{eqnarray} 
We would like to investigate thermodynamic of the  solutions using a redefinition of the $f(r)$ function in terms of a new function $F(r)$, similar to the procedure employed in \cite{Cai2009}. For instance, with the obtained solutions for $\lambda=1$  \eqref{lamda1fII} and general $\lambda$  \eqref{f2II} in mind,  defining 
\begin{eqnarray}\label{F}
	\begin{split}
		f(r)=-  \frac{4}{3}  {\frac {{r}^{2}
				\Lambda}{{{ n}}^{2}}}-  \frac{2}{3}  {\frac {{r}^{2-n}{\alpha}}{{{ n}}^{2
		}}}+{\frac {16  {r}^{2(1-n)}{{\alpha}}^{2} k_w^2}{9  M{{ n}}^{
					2}}}-F(r)	,
	\end{split}
\end{eqnarray} 
the Euclidean action takes the following considerably simplified form
\begin{eqnarray}\label{actiontII}
	\begin{split}
		I_E=&\frac {3\sqrt {2\alpha} {\kappa}^{2}{b^2}\beta\Omega}{32  k_w^4{M}^{2} \left( 4 
			\lambda -1\right) }\int dt dr  {{N}{r}^{\frac{3n}{2}-4}}\bigg[ {F}^{2}{M}^{2}{{n}}^{2} ( -3
		{{n}}^{2}+4 \lambda+6 {n}-4 )+{{
				n}}^{2}{{F'}}^{2}{M}^{2} \left( \lambda -1\right) {r}
		^{2} \\
		& +\frac {32}{81} {\alpha}^{2}{r}^{4-4 {n}} \left( 4 \lambda -1\right) 
		( 256 {\alpha}^{2} k_w^4+M ( -96 {
			r}^{{n}}\alpha k_w^2+9 M{r}^{2 {n
		}} )  )-r{{n}}^{2}{M}^{2}{
			F	F'}  \left( 3 {n}-4+4 \lambda \right)\bigg] 
		+B.
	\end{split}
\end{eqnarray} 
To have a well-defined variation principle the variation of the boundary term $B$  should have the following form 
\begin{eqnarray}\label{deltabII}
	\begin{split}
		\delta B&=\delta B_\infty-\delta B_{r_+}\\
		&={\frac {3 \sqrt {2\alpha}b^{2}{{ n}}^{2} {\kappa}^{2}\beta \Omega{} 
			}{32  \left( 4 \lambda-1
				\right) k_w^4 }}
		\big[  {r}^{ \frac{3}{2} { n}-3}  N 
		(  \left( 3 { n}-4+4 \lambda \right)F -2 r
		\left( \lambda -1\right)  {F'}
		) \delta F
		\big]^{\infty} _{r_+}.
	\end{split}
\end{eqnarray}
To evaluate this variation on the boundary at the horizon,  we will use the following identity for the variation of $F$ \cite{Banados}
\begin{equation}
	\delta F \mid _{r_+}=\bigg(\frac{\partial F}{\partial f}\bigg)_{r_+}[\delta f]_{r_+}, \\
\end{equation}
where 
\begin{equation}
	[\delta f]_{r_+}+\bigg(\frac{df}{dr}\bigg)_{r_+}\delta r_{+}=0,
\end{equation}
leads to
\begin{equation}\label{deltaF}
	\delta F \mid _{r_+}=-\bigg(\frac{\partial F}{\partial f}\bigg)_{r_+} \bigg(\frac{df}{dr}\bigg)_{r_+}\delta r_{+}=\bigg(\frac{df}{dr}\bigg)_{r_+}\delta r_{+}.
\end{equation}

\subsubsection{The $\lambda=1$ case}
In this case, the equations of motion of \eqref{actiontII} gives $N(r)$ and $F(r)$ in agreement with    \eqref{lamda1N} and \eqref{lamda1fII}, and we have 
\begin{equation}\label{FII1}
	\begin{split}
		F(r)=\frac{1}{9}{\frac {{r}^{2- n}}{Mb^8{{ n}}^{2}}}\big(& 9 {n} {b}^{8}{M}^{2} \left( 9 {C_1} {{n}}^{3}{b}^{8
		}+16 {a}^{4}\ln  \left( r \right)  \right) +256  \left( 3 {b}^{4}M-
		2 {a}^{2} k_w^2{r}^{-{n}} \right) {a}^{6} k_w^2{
			r}^{-{n}}
		\big) ^{\frac{1}{2}},
	\end{split}
\end{equation}
The constant $N_0$ in the lapse function \eqref{lamda1N} can be removed by a time redefinition and hence it is not a physical parameter. Also, the mass  $m$  and $N_0$ are a conjugate pair, where $N_0$ should be kept fixed while  $m$ is being varied \cite{Cai2009}. The only solution parameter that will be varied here is the $C_1$ constant, which is related to the physical parameter mass. 
Using \eqref{FII1}, at the boundary at infinity we have
\begin{equation}
	\delta B_{\infty}={\frac {3 \sqrt {2\alpha}\beta \kappa^2{{ n}}^{3}{b}^{2}}{64  k_w^4}}\Omega {N_0} \delta { C_1},
\end{equation}
and  on the horizon, using \eqref{deltaF},
the variation of the boundary term is given by
\begin{equation}
	\begin{split}
		\delta B_{r_+}= -{\frac {\sqrt {2
					\alpha}\pi \kappa^2{n} {b}^{2}\Omega  }{12Mk_w
				^{4}}}\big(& -8 {
			\alpha}^{2} k_w^2{r_+ }^{-{n}}+6 \Lambda 
		M{r_+ }^{{n}}+3 M\alpha \big){r_+ }^{\frac{n}{2}-1} \delta r_+.
	\end{split}
\end{equation}
Also, using \eqref{T1} and \eqref{T2},  
temperature in this class of solutions is given by
\begin{eqnarray}\label{}
	\begin{split}
		T=\frac {{N_0}   r_+^{\frac{3}{2}n}}{3{n} M\pi  }&\big( 44  \left( 8 r_+^
		{-{n}}\alpha k_w^2-3 M \right)k_w^2{\alpha}^
		{3}r_+^{-2 {n}}+12 r_+^{-{
				n}}{M}^{2}{\alpha}^{2}-3 {M}^{2}\Lambda  \left( 2 r_+^{{n}}\Lambda+\alpha \right)  \big)\\
		& \times\left( 6 r_+^
		{2 {n}}\Lambda M-8 {\alpha}^{2} k_w^2+3 Mr_+^{{n}}\alpha \right)^{-1}.
	\end{split}
\end{eqnarray}
Here, we can calculate the entropy either by using the $\delta B_+$ and free energy or by using the first law of thermodynamics (assuming its
validity)  $S=\int T^{-1}\frac{d m}{dr_+}dr_++S_0$, where $S_0$ is an integrating constant.\footnote{The $S_0$
	is an additive integrating constant that cannot be fixed using the first law of thermodynamics \cite{Kiritsis2010}. However, in some cases it can be selected appropriately based on some physical remarks \cite{Cai2009}.} Either way,  we get 
\begin{equation}\label{IIS}
	S ={\frac {\sqrt {2\alpha} b^{2}  {c}^{3}}{8MG\Lambda}}\left(2 \Lambda Mr_+^{\frac{3}{2}n}+3
	\alpha Mr_+^{\frac{n}{2}}+ 8 {\alpha}
	^{2} k_w^2r_+^{-\frac{n}{2}} \right)\Omega
	+S_0.
\end{equation} 
The first term is proportional to the area of the horizon. The entropy does not contain logarithmic correction term that is common in $(3+1)$ dimensional Ho\v{r}ava-Lifshitz black hole solutions,  but still diverges at $r_+\rightarrow 0$. The first two terms in the entropy resemble the entropy of $(4+1)$ dimensional black hole solutions with spherical and hyperbolic horizon \cite{Cai22009,Liu2013}. Here, for non-constant scalar curvature horizon with Nil geometry, the entropy included also the Ho\v{r}ava-Lifshitz parameter $k_w$ dependent term which is actually proportional to $A_H^{-\frac{1}{3}}$.

Also, using \eqref{d},  the mass recasts the following form in terms of the radius of horizon $r_+$
\begin{eqnarray}\label{}
	\begin{split}
		m=-&{\frac {\sqrt {2\alpha}\Omega b^{2}{c}^{3}{N_0} 
			}{48\pi n G
				\Lambda {M}^{2}}}	\big( 176 r_+^{-2 {n}}{\alpha}^{4}{{k_w}
		}^{4}-12 {\alpha}^{2}\ln  \left( r_+ \right) {n}
		{M}^{2}\\
		&-132 r_+^{-{n}}M{\alpha}^{3}{{k_w}
		}^{2}+3\Lambda{M}^{2} {r
			_+}^{{n}}( r_+^{ {n}}{\Lambda}+ \alpha) \big)+m_0,
	\end{split}
\end{eqnarray}
where $m_0$ is an integrating constant. 
To investigate  the local stability of the solutions we can consider the heat capacity, which using the mass and temperature is given by
\begin{eqnarray}\label{1C}
	\begin{split}
		C&=\frac{\partial m}{\partial T}\\
		&= -\frac {r_+^{\frac{3}{2}n} c^3 \sqrt {2
				\alpha}b^{2}\Omega}{24MG\Lambda} \left( 6 r+^{2 {n}}\Lambda M-8 {\alpha}^{2}{{k_w}}^{2
		}+3 Mr_+^{{n}}\alpha \right) ^{2}\\
		&\quad\times\bigg( - \frac{1}{2} {r
			_+}^{{n}}{\Lambda}^{2}{M}^{2}+\alpha  \big( {
			\frac {88}{3}} r_+^{-3 {n}}{\alpha}^{3}{{k_w}}^
		{4}-11 {\alpha}^{2} k_w^2r_+^{-2 {
				n}}M+r_+^{-{n}}{M}^{2}\alpha- \frac{1}{4} \Lambda
		{M}^{2} \big)  \bigg)\\
		&\quad\times \bigg( \frac{7}{3} M \left( 176k_w^2
		\Lambda-13 M \right) {\alpha}^{4} k_w^2r_+^
		{{n}}+{M}^{2}{\alpha}^{3} \left(M -112k_w^2\Lambda
		\right) r_+^{2 {n}}+{\frac {5 }{12}} \Lambda {M}^{
			2}{\alpha}^{2}r_+^{3 {n}}( 15M-16k_w^2\Lambda ) \\
		&\quad\quad\quad+r_+^{4 {n}}{\Lambda}^{2}{M}^{3}\alpha+r_+^{5 {n}}{
			\Lambda}^{3}{M}^{3}+\frac{176}{3} {\alpha}^{5}k_w^4  \left(3M - {4} r_+^{-{n}}\alpha k_w^2 \right)  \bigg).
		^{-1}
	\end{split}
\end{eqnarray}
It is not straightforward to use the heat capacity \eqref{1C} in its general form to determine whether this black hole solution is thermodynamic stable
or not. We would like, though, to provide some examples choosing particular set of values for the constants. For instance, with $\{n = 3, \Lambda = - 1/4, k_w = M=\alpha=b = 1\}$  the positive definiteness of  temperature demands  $r_+\gtrsim1.55$, where the heat capacity is always positive. On the other hand, for example with $\{n = 3, M = 1/3, k_w =  {1}/{2},\Lambda = -0.1, \alpha=b = 1\}$, temperature is positive definite between $1.54\lesssim r_+\lesssim 1.68$, where the heat capacity starting from zero, is positive until a divergent point at $r_+\approx1.6$ and then negatively approaches zero at the upper bound of $r_+$. Hence, similar to the other black hole solutions of Ho\v{r}ava-Lifshitz gravity \cite{Cai2009a}, depending on the values of parameters, this class of solutions can exhibit locally stable or non-stable behaviors.

As it has been mentioned before, in the $\lambda=1$ case  Ho\v{r}ava-Lifshitz gravity can reduce to general relativity. Black hole solutions for vacuum Einstein field equations with Nil geometry horizon have been obtained in \cite{Iizuka2012,PhysRevD.91.084054}, where suitable parameters have been selected to provide a horizon metric admitting additional isometry corresponding to Lifshitz scale invariance and hyperscaling violation. Although we have considered only the $3$-isometries of Bianchi types to obtain the solutions, the horizon metric in \eqref{metricII} can be rewritten in the form of the Lifshitz scale invariant metric given in \cite{Iizuka2012},  by setting $a={r_+^{-\frac{n}{2}}}$, or equivalently $\alpha=\frac{2}{3}r_+^{{ n}}\Lambda\, \left( 4p-3
	\right)
	$ in which for further simplicity we  set $b=\frac{3}{2\sqrt{-3\Lambda(3-4p)}}$, where $p$ is a constant. In this case, as $M\rightarrow \infty$  the  obtained thermodynamic quantities for $\lambda=1$ solutions behave  similar to the general relativity solutions with
	\begin{eqnarray}\label{}
		\begin{split}
			m={\frac {{c}^{3}r_{ +}^{\frac{5n}{2}}}{G\pi
			}}\Omega
			,\, S={\frac {{pc}^{3}}{G}}A_h,\, T=\frac {10}{3\pi p}\sqrt {-3\Lambda(3-4p)}r_+^{\frac{n}{2}}
			,\,\,\,
		\end{split}
	\end{eqnarray}
	where $A_h=\frac{3}{2\sqrt{-3\Lambda(3-4p)}}r_+^{2n}$ and $N_0$ has been eliminated by a rescaling in time. 
	The entropy in this limit, being propositional to the area of horizon, is in the form of Bekenstein-Hawking entropy. However, despite the Nil geometry solutions with hyperscaling violation \cite{PhysRevD.91.084054}, the entropy is positive with positive $T$.  

\subsubsection{A non-Einstein Case:  general $\lambda$ and $n=2$}
To investigate the thermodynamic behavior of this class of solutions, given by \eqref{n2N} and \eqref{n2ii},  noting that the $C_2$ constant in  \eqref{n2ii} can be removed by a rescaling of $r$, we define the new function $F(r)
$ by
\begin{eqnarray}\label{}
	\begin{split}
		f(r)=-\frac{\alpha}{6}+C_2 r^2-F(r),
	\end{split}
\end{eqnarray}
which yields the action
\begin{eqnarray}\label{I}
	\begin{split}
		I_E=
		\frac {\kappa^2}{k_w^{4}}\sqrt {2\alpha}b^{2}N\beta \Omega \int dtdr&\bigg( -{\frac {{\alpha}^{2}  }{9{M}^{2}
				{r}^{5}}}\big( 3k_w^2{r}^{
			3}M{ F'}-6 {r}^{2} k_w^2MF-3 {M}^{2}{r}^{4}+32 {r}^{2}{{
				k_w}}^{2}\alpha M-88  k_w^4{\alpha}^{2} \big)\\
		&+\frac {1}{8(4 \lambda-1)} \bigg( {3}{ { \left( \lambda-1 \right)   }}\left( {{F'}}^{2}{r}+4 {F}^{2}r^{-1} \right)
		+6 r\left( r{F'}+2 F \right)  \left( 3 {C_2}+\Lambda
		\right) \\
		&-6 \left( 2 \lambda+1 \right) F{F'}- 4{r}^{3}\left( 3 {
			C_2}+\Lambda \right) ^{2} \bigg)  \bigg) +B
		,
	\end{split}
\end{eqnarray}
whose equation of motion
gives 
\begin{eqnarray}
	F(r)=C_1 r^{-2},\quad N=N_0.
\end{eqnarray}
From the variation of \eqref{I}, we find that the variation of the boundary term $B$ should be given by   
\begin{eqnarray}\label{}
	\begin{split}
		\delta B&=-\frac {3\sqrt {2\alpha}b^{2}N\Omega \kappa^2
			\beta}{4M k_w^4 \left( 4 \lambda-1 \right) }\bigg[  \bigg( Mr
		\left(\lambda  -1\right) {F'}-M \left( 2 \lambda+1 \right) F+{r
		}^{2} M\left( \Lambda+3 {C_2} \right) -{\frac {4 k_w^2
				{\alpha}^{2}}{9{r}^{2}}}\left( 4 \lambda -1\right) \bigg)\delta F \bigg]^{\infty} _{r_+}.
	\end{split}
\end{eqnarray}
On the horizon  using $f(r_{+})=0$ and \eqref{deltaF} we obtain
\begin{eqnarray}\label{}
	\begin{split}
		\delta B_{r_+}= &{\frac {-b^{2}\kappa^2\sqrt {2\alpha}\pi \Omega 
			}{18r_+^{2}M k_w^4 \left( 
				4 \lambda-1 \right) }}\big( 54 M \left( \Lambda-4 { C_2}  \left( \lambda+1 \right) 			\right) r_+^{4}+3  \alpha\left( 4 \lambda -1\right) 
		\left( 3 r_+^{2}M-8 \alpha {{k_w}}^{2
		} \right)  \big)
		\delta r_{+},
	\end{split}
\end{eqnarray}
and at infinity  we have
\begin{equation}\label{dbinf22}
	\delta B_{\infty}=-{\frac {3\beta\kappa^2b^{2}\sqrt {2\alpha} \Omega }{4{{k_w
			}}^{4} \left( 4 \lambda -1\right) }}\left( \Lambda+3 {C_2} \right)N_0	 \delta { C_1}.
\end{equation}
Removing the variations from this kind of equations to obtain the mass and entropy needs boundary conditions to be imposed \cite{Hertog,Anabal2015}.\footnote{{In fact, it was first observed  in  AdS context in \cite{Hertog} that for Einstein-Scalar models with scalar filed $\phi=\frac{\alpha}{r}+\frac{\beta}{r}$  the integrability of  energy in Hamiltonian formalism, which contains $\delta Q_{\phi}=\int \beta \delta \alpha d\Omega+\dots$ term, forces  the coefficients  in the asymptotic expansion of the scalar field  $\alpha$ and $\beta$   to be functionally related.}} In particular, it requires  $C_2$ to be functionally related to
$C_1$. 
We will take advantageous of the first law of thermodynamics to determine this functional relation. 

The temperature of the black hole can be computed using Euclidean regularity, which gives
\begin{eqnarray}
	\begin{split}
		T={\frac {N_0  }{12\pi r_+ }}\left( 12 {C_2}r_+ ^2-\alpha \right).
	\end{split}
\end{eqnarray}
The first law of thermodynamics
\begin{equation}
	dm=TdS,
\end{equation}
is then satisfied if the $C_2$ constant takes one of the following forms
\begin{eqnarray}\label{c22ii}
	\begin{split}
		C_2^{{\rm ex}}=\frac{\alpha}{12 r_+^2}
		,\quad C_2={\frac {\alpha \left( 3 M{r_+ }^{2}-8 \alpha
				k_w^2 \right) }{18M{r_+ }^{4}}}
		,
	\end{split}
\end{eqnarray}
which also gives the relation between $C_1$ and $C_2$ using $f(r_+)=0$.
The first expression, being denoted by the "ex" symbol that here and hereafter  stands for the extremal case,  coincides with the determined $C_2$ by the condition of degenerate horizon   $f(r)=f'(r)=0$
. Generally, the $C_2$ parameters in \eqref{c22ii} are accompanied by the following $C_1$ expression, respectively, 

\begin{eqnarray}
	\begin{split}
		C_1^{{\rm ex}}=-\frac{\alpha r_+^2}{12},
		\quad C_1=-\frac{4k_w^2\alpha^2}{9M}.
	\end{split}
\end{eqnarray}
Considering these points, the extremal radius of horizon is given by
\begin{eqnarray}\label{Mexii}
	\begin{split}
		r_+^{{\rm ex}}=\frac{4\sqrt{3\alpha M} k_w}{3M}.
	\end{split}
\end{eqnarray}
Now, performing the integrals,   the mass and entropy are obtained as follows
\begin{eqnarray}\label{}
	\begin{split}
		m=&\frac {{c}^{3}b^{2}{\alpha}^{ \frac{3}{2}}\sqrt {2}{N_0} \Omega}{96 
			\Lambda G\pi {M}^{2}} \big( Mr_+^{4}\alpha  \left(3M -32k_w^2\Lambda \right) 
		\ln  \left( r_+ \right)r_+^{4} +3 \Lambda {M}^{2}r_+^{6}+12 {\alpha}^{2} k_w^2Mr_+^{
			2}-{\frac {32 {\alpha}^{3} k_w^4}{3}} \big) 
		+m_0,
	\end{split}
\end{eqnarray}
\begin{eqnarray}\label{}
	\begin{split}
		S= {\frac {\sqrt {2\alpha}b^{2} {c}^{3}\Omega}{8 M\Lambda G}}
		\left(2 M\Lambda r_+^{3}+3 M
		\alpha r_++8k_w^2{\alpha}^{2}r_+^{-1}
		\right)
		+S_0.\quad
	\end{split}
\end{eqnarray}
Also, the heat capacity is given by
\begin{eqnarray}\label{Cn2II}
	\begin{split}
		C=- \frac{b^
			{2} \sqrt {2\alpha}{c}^{3}\Omega}{24 MG
			\Lambda } &\frac {  \left( 3 Mr_+^{2}-16 
			\alpha k_w^2 \right) }{r_+ \left( Mr_+^{2}-16 \alpha k_w^2
			\right) }\left( 6 Mr_+^{4}\Lambda+3 M{r_{
				\mbox {{\tt +}}}}^{2}\alpha-8k_w^2{\alpha}^{2} \right).
	\end{split}
\end{eqnarray}
Evidently, the thermodynamic behavior is independent of the value of $\lambda$. Generally, the heat capacity vanishes when $r_+$ equals to one of the following radii
\begin{eqnarray}\label{}
	\begin{split}
		&	r_1^2= {\frac {16\alpha k_w^2}{3M}},\\
		& r_{2,3}^2=- {\frac {  \alpha}{12M\Lambda}}\left( 3 M\mp\sqrt {192 M\Lambda k_w^2+9 {M
			}^{2}} \right),
	\end{split}
\end{eqnarray}
and diverges when
\begin{eqnarray}\label{}
	\begin{split}
		r_+=\sqrt{3}r_1\equiv r_4
	\end{split}
\end{eqnarray}
Temperature is positive definite for $r_+\geqslant r_1$ and  $r_1$ is actually the extremal point indicated by \eqref{Mexii}.  Obviously, the heat capacity divergent point $r_4$ is in the positive temperature range of $r_+$. But for the zero points $r_2$ and $r_3$, depending on the values of  parameters, we can have different situations:

(i) If a set of parameters is selected for which   ${192 \Lambda M k_w^2+9 {M}^{2}}<0$,  there are no real $r_2$ and $r_3$ and the solutions at $r_+\geqslant r_1$ range are stable until $r_+=r_4$ and then become unstable.\footnote{For example with $\{\Lambda = - \frac{1}{4}, M =  \frac{1}{2}, k_w =  \frac{1}{2}, \alpha = 1\}$. }

On the other hand,  if  ${192 \Lambda M k_w^2+9 {M}^{2}}>0$, the $r_2$ and $r_3$ are both real. Then, 

(ii) If one selects a set of parameters for which $r_2<r_1<r_4<r_3$, then  the solutions are unstable in $r_1<r_+<r_4$, stable  after the divergent point $r_4$  in $r_4<r_+<r_3$ range, and then become unstable.\footnote{For example with $\{\Lambda = -0.1,  M = 4, k_w = \alpha = 1\}$.}

(iii)
If   a set of parameters is selected to  have  $r_2<r_1<r_3<r_4$, the solutions with $T>0$ are unstable as $r_1<r_+<r_3$,  stable as $r_3<r_+<r_4$, and then become unstable.\footnote{For example with $\{\Lambda = - \frac{1}{2}, M =  \frac{1}{2}, k_w = 0.6, \alpha = 1\}$. }

\subsubsection{Non-Einstein Case:  general $\lambda$}\label{generallII}
When the Ho\v{r}ava-Lifshitz coupling constant $\lambda$ is allowed to have any value, the black hole solutions with Nil geometry horizon have been obtained as given by \eqref{N2II} and \eqref{f2II}. As we have mentioned earlier, we intend to consider two cases in this class of solutions. First, in the most general case when $n$ and $\lambda$ have general and primarily independent values,  noting the asymptotic behavior of metric, similar to the solutions obtained in \cite{Cai2009}, our emphasis is on the negative power of $r$ branch in \eqref{f2II},  indicated by $C_1$ term.  The second case is
when there is a special relation of type \eqref{lanadII} between $\lambda$ and  $n$, which leaded to $f(r)$ function of the form \eqref{sol1II}. 

In the first category, for which only the $C_1$ term in the $f(r)$ function  \eqref{sol1II} is present, to investigate the thermodynamics in Hamiltonian formalism we consider the $F(r)$ function defined by \eqref{F}, where the Euclidean action takes the form of \eqref{actiontII}, and the solution for its equations of motion  
gives
\begin{eqnarray}\label{Fl}
	F(r)=C_1 r^{s},\quad N=N_0, \quad s=-\frac{3}{4}(n-2)-\sqrt{\mu},
\end{eqnarray}
where $\mu$ is given by \eqref{mu}. Noting the variation of the boundary term of Euclidean action, given by \eqref{deltabII}, it is easy to find that a definite and non-vanishing   $\delta B_{\infty}$ demands
\begin{eqnarray}\label{con1}
	2s+\frac{3}{2}n-3=0,
\end{eqnarray}
which, practically, relates the $n$ constant in metric to the $\lambda$ parameter by 
\begin{eqnarray}\label{con2}
	\lambda={\frac {21 {{n}}^{2}-12 {n}+4}{9 {{n}}^{2}+12 {n}+4}}.
\end{eqnarray}
This is, on the other hand, equivalent to $\mu=0$, which practically leaves no difference between $C_1$ and $C_2$ dependent terms in \eqref{actiontII}.
Applying this condition we get
\begin{eqnarray}
	\delta B_{\infty}={\frac {3  \beta {\kappa}^{2}{{n}}^{3}{b}^{2}{
				N_0}\left( 3n+2 \right) }{ 32\left( 5 {n}-2 \right)  k_w^4}}\sqrt {2\alpha}{ C_1} \delta { C_1} \Omega,
\end{eqnarray}
\begin{eqnarray}
	\begin{split}
		\delta B_{+}=-&{\frac {\sqrt {2\alpha}\pi {n} {b}^{2} {\kappa}^{2}
				\left( 3 {n}+2 \right)  }{12M \left( 5 {n}-2 \right)  k_w^4}} \big( -8 {r_+
		}^{-2 {n}}{\alpha}^{2} k_w^2+6 \Lambda M+3 r_+^{-{n}}\alpha M \big)r_+^{\frac{3}{2} {n}-1}\Omega.
	\end{split}
\end{eqnarray}
Also, temperature of this black hole solution is given by
\begin{eqnarray}
	\begin{split}
		T={\frac {{N_0}  r_+}{
				72 M{{n}}^{2}\pi}} \big(& 8 {\alpha}^{2} k_w^2 \left( 2-5 {
			n}\right) r_+^{-2 {n}}+3 \alpha M
		\left( n-2 \right) r_+^{-{n}}-6 
		\Lambda M \left( 3 {n}+2 \right)  \big).
	\end{split}
\end{eqnarray}
These thermodynamic quantities satisfy the first law of thermodynamics. Also, the $C_1$ constant can be written in terms of radius of horizon as follows
\begin{eqnarray}
	C_1=-{\frac { 2 r_+^{\frac{3}{4}{n}+ \frac{1}{2}}}{9M{{n
			}}^{2}}}\left( -8 r_+^{-2 {n}}{\alpha}
	^{2} k_w^2+6 \Lambda M+3 r_+^{-{n}}
	\alpha M \right).
\end{eqnarray}
Noting this, we can perform the integrating to obtain the mass and entropy as follows
\begin{eqnarray}
	\begin{split}
		m=-\frac { \Omega b^{2} {N_0} \sqrt {2\alpha}{c}^{3}
			\left( 5 {n}-2 \right) }{576 \pi  \left( 3 {n}+2
			\right) G\Lambda {M}^{2}{n}}r_+^{\frac{3}{2}n+1} \big( &-16 {\alpha}^
		{2} k_w^2r_+^{-2 {n}} \big( 6 \Lambda M-4 r_+^{-2 {n}}{\alpha}^{2} k_w^2
		+3 r_+^{-{n}}\alpha M \big) \\
		&+9 {M}^{
			2} \left( 4 {\Lambda}^{2}+4 \Lambda r_+^{-{n
		}}\alpha+{\alpha}^{2}r_+^{-2 {n}} \right) 
		\big)+m_0
		,
	\end{split}
\end{eqnarray}
\begin{eqnarray}
	\begin{split}
		S= \frac {{c}^{3} \left( 5 {n}-2 \right) b^{2}\Omega 
			\sqrt {\alpha}   }{8
			\Lambda G \left( 3 {n}+2 \right) M}\big(& 2 \Lambda Mr_+^{ \frac{3}{2} {
				n}}+8 {2\alpha}^{2} k_w^2r_+^{- \frac{n}{2}}+3 \alpha Mr_+^{ \frac{n}{2}} \big)
		+S_0.
	\end{split}
\end{eqnarray}
The first term in entropy is proportional to the area of horizon $A_H$, given by \eqref{Ahii}, and there is a divergent in mass and entropy at $r_+\rightarrow 0$ limit.
The heat capacity for this class of solution is given by
\begin{eqnarray}
	\begin{split}
		C=&-\frac {9 b^{2}{{  
					n}}^{3}\Omega {c}^{3}\sqrt {2\alpha}   \left( 5 {n}-2
			\right)   }{32 \Lambda G \left( 3 {n}+2 \right) 
		}r_+^{\frac{3}{4} {n}- \frac{1}{2}}{C_1}\big( 8 {\alpha}^{2} k_w^2 \left( 5 {n}-2
		\right) r_+^{-2 {n}}-3 M ( \alpha 
		\left( n-2 \right) r_+^{-{n}}-2 
		\Lambda  \left( 3 {n}+2 \right)  )  \big)\\
		&\times\big( -8 {\alpha}^{2} \left( 2n-1 \right)  \left( 5 {
			n}-2 \right)k_w^2r_+^{-2 {n}}+3
		M \left( \alpha  \left( n-1 \right)  \left( n-2
		\right) r_+^{-{n}}+2 \Lambda  \left( 3 {
			n}+2 \right)  \right)  \big)^{-1}
		,
	\end{split}
\end{eqnarray}
which vanishes when $r_+$ equals to either of the following radii
\begin{eqnarray}\label{r0}
	\begin{split}
		& r_{1,2}^{2n}=-{\frac {\alpha}{12
				\Lambda Mb^{2}}}\left(3 M\mp\sqrt {192 \Lambda M k_w^2+9 {M}^{2}}\right),
		\\
		&	r_{3,4}^{2n}= \frac {1}{16 k_w^2
			\left( 5 {n}-2 \right) \alpha}\big(3 M \left( n-2 \right)\mp\big(9  \left( {
			n} -2\right) ^{2}{M}^{2}-192 M\Lambda k_w^2 \left( 3 {n}+2
		\right)   \left( 5 {n}-2 \right) \big)^{\frac{1}{2}}\big),
	\end{split}
\end{eqnarray}
and diverges when $r_+$ equals to the following radii
\begin{eqnarray}\label{}
	\begin{split}
		r_{5,6}^{-2n}=& \frac {1}{16 k_w^2 \left( 10 {{n}}^{2}-9 
			{n}+2 \right) \alpha}\big(3 M \left( {{n}}^{2}-3 {n}+2 \right)\\
		& \mp\big(9 {M}^{2} \left( {{n}}^{2}-3 {n}+2 \right) ^{2}+192 
		\Lambda M k_w^2 ( 30 {{n}}^{3}-7 {{n}}^{2}-
		12 {n}+4 ) \big)^{\frac{1}{2}}\big)
		.
	\end{split}
\end{eqnarray}
Noting $\Lambda<0$, $r_3$ is not real, even for odd valued $n$.
As  $r_+=r_4$, this category of black hole solutions becomes extremal with
vanishing temperature. The reality of these radii and local stability  depend on the values of parameters. For example, choosing the set of parameters $\{n = 4, \Lambda = -0.02, k_w=M= \alpha = 1\}$ that keeps only the $r_1<r_4<r_2$  real,  the temperature is positive as $r_+>r_4$ and in this region the heat capacity is negative until  $r_+=r_2$, and then becomes positive without any divergent. As an another example, if one sets  $\{n=4,M=2,\Lambda=-0.01,k_w=0.1,\alpha=1\}$, the real radii are in the order of  $r_1<r_4<r_6<r_5<r_2$. Here, the positive definiteness of temperature demands, again, $r_+>r_4$.  In this region, the solutions are unstable as $r_4<r_+<r_5$, then become stable between two divergent points $r_5$ and $r_6$, and then the unstable phase in $r_5<r_+<r_2$ range is followed by another stable phase where $r_+>r_2$.

In the second case of this class of solutions we consider
the category of parameters for the solutions \eqref{f2II}  in which  $\lambda$ and $n$ are related to each other by the relation \eqref{lanadII}, and consequently the $f(r)$ function reduces to \eqref{sol1II}, in which no asymptotic  problem occurs in presence of the $C_2$ constant. 
As we have mentioned before,  the  $\lambda=1$ case in this class of solutions, which corresponds to $n=2$, is similar to the solutions for $\lambda=1$  presented in \cite{Liu2013}, however with different horizon geometries of flat, spherical, and hyperbolic. 
Our focus here is on $n>2$ case.  
The solution contains two integrating constant $C_1$ and $C_2$, while the only physical parameter characterizing this black hole is mass. 
In topological black hole solutions, whose horizons are constant curvature Einstein spaces, the requirement of asymptotic AdS behavior of the spacetime relates the integrating constant of type $C_2$  to the curvature constant $k$ of the horizon \cite{Topologicalbh1}. However,  the Bianchi type $II$ space does not admit  Einstein space metric and the horizon curvature depends on $r_H$. Hence, we consider $C_2$  as a yet undetermined constant.
Then, rewriting the action in terms of the new function $F(r)$ defined  by 
\begin{eqnarray}\label{}
	\begin{split}
		F(r)=-  \frac{4}{3}  {\frac {{r}^{2}
				\Lambda}{{{ n}}^{2}}}-  \frac{2}{3}  {\frac {{r}^{2-n}{\alpha}}{{{ n}}^{2
		}}}&+{\frac {16  {r}^{2(1-n)}{{\alpha}}^{2} k_w^2}{9  M{{ n}}^{
					2}}}+C_2-f(r)	,
	\end{split}
\end{eqnarray} 
we obtain the Euclidean action 
\begin{eqnarray}\label{action3}
	\begin{split}
		I_E=-\beta\Omega\int& dr \frac {3 {r}^{ \frac{3}{2} { n}}{b}^{2}\sqrt {2\alpha}{k
			}^{2} N  }{32 {M}^{2} k_w^4 \left( {
				n}+2 \right) } \bigg({\frac {{{ n}}^{3}  {M}^{2} \left( 2-n
				\right)F'^{2} }{4{r}^{2} \left( { n}-1 \right) }}+ 3 { {{M}^{2}{{ n}}^{3} \left( n-2 \right)  \left(F -{ C_2} \right) ^{2}}{{r
				}^{-4}}}\\
		&+4 { {{M}^{2}{{ n}}^{3} F' \left( F  -{ C_2}
				\right) }{{r}^{-3}}}-{\frac { 32 {\alpha}^{2}  }{81}}\left( {
			n}+2 \right)( 256 {r}^{-4 { n}}{\alpha}
		^{2} k_w^4-96 M{r}^{-3 { n}}\alpha  k_w^2+9 {M
		}^{2}{r}^{-2 { n}} ) \bigg) +B
		,
	\end{split}
\end{eqnarray}  
whose equations of motion gives
\begin{eqnarray}\label{1F}
	F(r)=C_1 r^{-\frac{3n}{2}+3},\quad N=N_0.
\end{eqnarray}
The variation of the boundary term for  \eqref{action3} should be given by   
\begin{eqnarray}\label{}
	\begin{split}
		\delta B=&{\frac {3 \sqrt {2\alpha} \beta 
				\kappa^2{b}^{2}{{ n}}^{3}\Omega }{ 64\left( { n}+2
				\right)  \left( { n}-1 \right)  k_w^4}}\big[N{r}^{ \frac{3}{2} { n}-3} \big( 
		\left(2-n  \right)  r{ F'}+8 \left( { n}-1 \right) 
		\left( F-{ C_2} \right)  \big)\delta F 
		\big]^{\infty} _{r_+}.
	\end{split}
\end{eqnarray}
Then, using \eqref{1F}, at infinity and on the horizon we obtain 
\begin{equation}\label{dbinfII}
	\delta B_{\infty}=-{\frac {3\beta \sqrt {2\alpha}N_0  {{ n}}^{3}\kappa^2{b}^{
				2}\Omega}{8 k_w^4 \left( { n}+2 \right) }}{ C_2}\delta { C_1}
	,
\end{equation}
\begin{eqnarray}\label{dbinfII2}
	\begin{split}
		\delta B_{r_+}=-\frac {\kappa^2 \pi n \sqrt {2\alpha
			}{b}^{2} r_+^{ \frac{3}{2} { n}}\Omega}{96 M \left( { n}+2 \right) 
			\left( { n}-1 \right)  k_w^4} \big(& 2 \left( { n}+
		2 \right) \left( 3 { n}-2 \right)  r_+^{-2 { n}-1} ( -8
		{\alpha}^{2} k_w^2+6 r_+^{2 { n}}\Lambda M+3 r_+^{{
				n}}\alpha M ) \\
			&-27 { {{ C_2} {{ n}}^{2}
				\left( n-2 \right) ^{2}M}{ r_+^{-3}}} \big)  \delta r_+.
	\end{split}
\end{eqnarray}
Similar to \eqref{dbinf22}, removing the variations from these  equations to obtain the mass and entropy needs boundary conditions to be imposed as a functional relation between $C_2$  and
$C_1$, or equivalently $r_+$. 
To obtain the explicit form of this functional relation we establish the first law of thermodynamics. Noting that, based on Euclidean regularity, the temperature  is given  by
\begin{eqnarray}
	\begin{split}
		T=&{\frac {N_0  }{72 rM{{ n}}^{2}\pi}}\big( -3 ( 2 \alpha \left( n-2	\right)  r_+^{2-{ n}}+4 r_+^{2}\Lambda \left( 3 { n}-2		\right) -9 { C_2} {{ n}}^{2} \left( { n} -2\right) 	) M-16  k_w^2{\alpha}^{2} \left( { n}+2	\right)  r_+^{2-2 { n}} \big),
	\end{split}
\end{eqnarray}
the  first law of thermodynamics 
revels that the obtained thermodynamic quantities satisfy the first law
if we have
\begin{eqnarray}\label{c2II}
	\begin{split}
		C_2=\frac {r_+^{2}  }{27 \left( {n
			} -2\right) {{n}}^{2}M}
		\big(& -16k_w^2{\alpha}^{2} \left( {n}-4
		\right) r_+^{-2 {n}}+12 M \left( \alpha 
		\left( n-2 \right) r_+^{-{n}}+\Lambda
		\left( 3 {n}-4 \right)  \right)\\
		& \pm 2 {r_+
		}^{-2 {n}}{  }  ( -12 \Lambda Mr_+
		^{2 {n}}+3 M\alpha  \left( n-2 \right)r_+^{{n}}-16k_w^2{\alpha}^{2} \left( {n}-1
		\right)  )  \big),
	\end{split}
\end{eqnarray}
which also fixes the relation between $C_1$ and $C_2$ using $f(r_+)=0$. Substituting the expression of $C_2$ with positive sign in \eqref{c2II} into $f(r_+)=0$ leads to $C_1r_+^{\frac{3}{2}(n-2)}=0$, which is not an interesting case. On the other hand, the negative sign in \eqref{c2II} gives an expression for  $C_2$ coincided with the $C_2$ obtained in extremal case from the conditions $f(r_+)=f'(r_+)=0$, which is accompanied by the following form of $C_1$ constant
\begin{eqnarray}\label{extermcIIc1}
	\begin{split}
		C_1^{{\rm ex}}
		= \frac{1}{27}&{\frac { r_+^{\frac{3}{2}n-1}}{
				\left( n-2 \right) {{n}}^{2}M}} \big( 64k_w^2{\alpha}^{2} \left( {n}-1
		\right) r_+^{-2 {n}}-12 M \left( \alpha 
		\left( n-2 \right) r_+^{-{n}}-4 
		\Lambda \right)\big).
	\end{split}
\end{eqnarray}
Hence, the  consistent solutions in this class is the extremal case.  Temperature vanishes for this solutions and entropy is given by
\begin{eqnarray}\label{}
	\begin{split}
		S^{{\rm ex}}=&\frac {b^{2}{c}^{3}  \sqrt {2\alpha}\Omega 
		}{16M	G\Lambda \left( 3 {n}-2 \right) 
		}\big(4M \Lambda 
		\left( 3 {n}-2 \right)r_+^{\frac{3}{2}n}+ 16 {\alpha}^{2}{{k_w}
		}^{2} \left( {n}+2 \right)  \left( {n}-1 \right) r_+^{-\frac{n}{2}}+3M  \alpha  \left( {{n}}^{2}+4
		{n}-4 \right) r_+^{\frac{n}{2}}    \big).
	\end{split}
\end{eqnarray}
The near  horizon geometry for this extremal solutions can be obtained by using the following  change of the variables
\begin{eqnarray}\label{}
	\begin{split}
		r\rightarrow \left(r_+^{2-2n}+\frac{\epsilon}{r}\right)^{\frac{1}{2-2n}},\quad t\rightarrow \frac{t}{\epsilon},
	\end{split}
\end{eqnarray}
and then sending $\epsilon\rightarrow 0$,  which yields the near horizon metric as follows
\begin{eqnarray}\label{}
	\begin{split}
		ds^2=-N_0 W\frac{dt^2}{r^2}&+\frac{1}{W}\frac{dr^2}{r^2}+r_+^n\bigg(a \left(dx^2-x^1dx^3\right)^2+b \left(\left(dx^1\right)^2+\left(dx^3\right)^2\right)\bigg),
	\end{split}
\end{eqnarray}
where
\begin{eqnarray}\label{}
	\begin{split}
		W=&{\frac {  r_+^{2 {n}-2}}{72 M{{n}}^{2}
				\left( {n}-1 \right) ^{2}}}\big( 16 {\alpha}^{2} k_w^2 \left( {n}+2
		\right)  \left( {n}-1 \right) + 3M\alpha  \left( n-2 \right) ^{2}r_+^{{n}}-12  \left( 3 {n}-2 \right)M \Lambda {r_+}^{2 {n}}
		\big)
		.
	\end{split}
\end{eqnarray}
By a scaling of time, the above solution can be rewritten as a product space of $AdS_2 \times Nil$ 
with different radii
\begin{eqnarray}\label{}
	\begin{split}
		ds^2=\frac{1}{W}& \frac{-dt^2+dr^2}{r^2}+r_+^n\bigg(a \left(dx^2-x^1dx^3\right)^2+b \left(\left(dx^1\right)^2+\left(dx^3\right)^2\right)\bigg).
	\end{split}
\end{eqnarray}

\subsection{Thermodynamics of Bianchi type III solutions with $H^2\times R$ horizon geometry}
Thurston closed geometries of  product  constant curvature  type $H^2\times R$  and  twisted product type $ \widetilde{	{SL_2R}}$ can locally possess  Bianchi type $III$ symmetry
with  $ SO(2)$ isotropy \cite{closedBianchi}. In our considered case, for the Bianchi type $III$ symmetric spacetime metric \eqref{metricIII} the  geometry of horizon is equivalent to $H^2\times R$.
Families of black hole solutions for this Bianchi type have been obtained in \eqref{n1}-\eqref{f2} for $\lambda=1$ and general $\lambda$, given in terms of horizon curvature related parameter $\alpha=\frac{2}{a}$.
The area of horizon for these solutions is  given by
\begin{eqnarray}
	A_H
	=\frac{\sqrt{b}}{\alpha}r_+^{\frac{3n}{2}}.
\end{eqnarray}
Having found the solutions for $f(r)$ function beforehand, to investigate the thermodynamics of the solutions \eqref{lamda1f} and \eqref{f2}   we can rewrite the action in terms of new $F(r)$ function
\begin{eqnarray}\label{Fiii}
	\begin{split}
		F(r)=-\frac{4}{3}\frac{r^2 \Lambda}{n^2}-\frac{2}{3}\frac{\alpha r^{2-n}}{n^2 }+\frac{1}{9}\frac{ \alpha^2 r^{2-2 n} k_w^2}{n^2 M}-f(r)	,
	\end{split}
\end{eqnarray} 
which gives the Euclidean action
\begin{eqnarray}\label{actiontIII}
	\begin{split}
		I_E=-\beta\Omega\int dtdr\frac {9\sqrt {b} {r}^{-4+\frac{3}{2} n}{\kappa}^{2}N  }{16 {M}
			^{2}\alpha  k_w^4 \left( 4 \lambda-1 \right) } \bigg[&\frac{{M}^{2}}{3}\left(1- \lambda \right)
		( {{F'}}^{2}{r}^{2}-4F {F'}r+4 {F}^{2} ) {n}^{
			2}+ {n}^{
			3} \left( F \left( n-2 \right) +r{F'} \right) F{M}^{2}\\
		& -{\frac {2     {\alpha}
				^{2}}{243}}\left( 
		4 \lambda-1 \right)( 3 M{r}^{n}
		\left(3 M{r}^{n}  -2k_w^2\alpha\right)  + k_w^4{\alpha}^{2}){r}^{4-4 n} \bigg] 
		+B.
	\end{split}
\end{eqnarray}
Here, the variation of boundary term $B$ must have the following form
\begin{eqnarray}\label{boundariiii}
	\begin{split}
		\delta B
		=&{\frac {3{\kappa}^{2}\sqrt {b} {n}^{2}\beta\Omega}{16\alpha {{k_w}
				}^{4} \left( 4 \lambda-1 \right) }}
		\left[\left( 3  nF+2  \left( \lambda-1 \right)  \left( 2
		F-{ F'} \right)  \right)   N
		{r}^{\frac{3}{2} n-3}\delta F 
		\right]^{\infty} _{r_+}.
	\end{split}
\end{eqnarray}
The solutions for this Bianchi type have similar structure to those of Bianchi type $II$, and to study the thermodynamic properties we will follow the same procedures here.

\subsubsection{The $\lambda=1$ case}
With $\lambda=1$, the solution of equations of motion of action \eqref{actiontIII}, being in agreement with   \eqref{n1}  and \eqref{lamda1f}, gives
\begin{equation}\label{2F}
	\begin{split}
		F(r)=\frac{r^{-n+2}}{9n^2 M}& \big(2{\alpha}^{3}{{
				k_w}}^{2}{r}^{-n} \left( 12M - k_w^2{r}^{-n}
		\alpha\right) +9{M}^{2}n \left( 9{ C_1}{n}^{3}+4\ln 
		\left( r \right) {\alpha}^{2} \right) 
		\big)^{\frac{1}{2}} .
	\end{split}
\end{equation}
Using $f(r_+)=0$, temperature in this class of solutions is given by
\begin{eqnarray}\label{}
	\begin{split}
		T=\frac {{N_0} r_+^{ \frac{n}{2}}  }{12Mn\pi  }&\big(  \left(  k_w^2r_+^{-n	}\alpha-6 M \right) r_+^{-n}{\alpha}^{3} k_w^2+6 {M}^{2}	\left( -8 {\Lambda}^{2}r_+^{2 n}+\alpha \left( -4 \Lambda r_+^{	n}+\alpha \right)  \right)  \big)\\
		&\times\left( 12 Mr_+^{2 n}\Lambda	- k_w^2{\alpha}^{2}+6 Mr_+^{n}\alpha \right)^{-1}.
	\end{split}
\end{eqnarray}
Also, using \eqref{boundariiii} and \eqref{2F} we have
\begin{equation}
	\delta B_{\infty}=\frac {3  \sqrt {b}\beta {n}^{3}\kappa^2\Omega}{32\alpha  k_w^{4}}{N_0} \delta {
		C_1},
\end{equation}
\begin{equation}
	\begin{split}
		\delta B_{r_+}= -{\frac {  r_+^{-\frac{n}{2}-1}n\sqrt {b}{
					k}^{2}\pi \Omega}{12\alpha  k_w^4M}}\big(& 12 Mr_+^{2 n}\Lambda- k_w^2{\alpha}^{2}+6 Mr_+^{n}\alpha \big)
		\delta r_+.
	\end{split}
\end{equation}
The first law of thermodynamics is satisfied and the mass and  entropy can be calculated up to  additive constants as follows
\begin{eqnarray}\label{}
	\begin{split}
		m=&
		-{\frac {\sqrt {b}{c}^{3}\Omega   {N_0}	}{384n\alpha\pi{M}^{2} G\Lambda}}\big(k_w^{2}{\alpha}^{3}{r_{
				\mbox {{\tt +}}}}^{-n} \left( r_+^{-n} k_w^2
		\alpha-12 M \right)+12 {M}^{2} \left( -n{\alpha}^{2}\ln  \left( r_+
		\right) +4 {\Lambda}^{2}r_+^{2 n}+4 \Lambda {r_
			+}^{n}\alpha \right)  \big)
		+m_0
	\end{split}
\end{eqnarray}
\begin{equation}\label{}
	S =
	{\frac { \sqrt {b}{c}^{3} \Omega }{8\alpha MG		\Lambda}}\left(4 \Lambda M r_+^{\frac{3n}{2}}+6 M\alpha r_+^{\frac{n}{2}}+ {\alpha}^{2} k_w^2r_+^{-\frac{n}{2}}
	\right)
	+S_0.
\end{equation} 
Also, the heat capacity is given by
\begin{equation}\label{}
	\begin{split}
		C =&\frac {\sqrt {b}\Omega
			^{2}{c}^{3}r_+
			^{ -\frac{3}{2} n}}{4608  \alpha MG\Lambda}  \left( 12 \Lambda Mr_+^{2 n}-{
			\alpha}^{2} k_w^2+6 Mr_+^{n}\alpha \right) \big(  \left(6 M -r_+^{-n}
		\alpha k_w^2 \right) r_+^{-n}{\alpha}
		^{3} k_w^2+6 {M}^{2} ( 8 r_+^{2 n}{
			\Lambda}^{2}+4 \Lambda r_+^{n}\alpha-{\alpha}^{2}
		)  \big)\\
		&\times \bigg( {\frac {{\alpha}^{4} 
			}{96}}\left( 14 {{k_w
		}}^{2}\Lambda-17 M \right)k_w^2Mr_+^{-n}+\frac{1}{16} {M}^{2}{\alpha}^{3} \left( M-12k_w^2\Lambda
		\right)+{\Lambda}^
		{2}{M}^{3}r_+^{2 n}\alpha \\
		&\quad\quad+{\frac {5 }{24}}{\alpha}^{2}\Lambda {M}^{2}r_+^{n} \left(3M -2k_w^2
		\Lambda \right) +{\Lambda}^{3}{M}^{3}{r_+}^{3 n}+\frac{1}{16} Mr_+^{-2 n}{\alpha}^
		{5} k_w^4-{\frac {r_+^{-3 n}{\alpha}^{6}{{
						k_w}}^{6}}{192}} \bigg) ^{-1}.
	\end{split}
\end{equation}
Depending on the values of parameters we can have stable and unstable solutions here. 
For example with $\{\Lambda = -1/3, k_w = 2,  n = 3,\alpha =  M = 1\}$,  temperature is positive definite at $r_+\gtrsim1.02$, where the heat capacity is always positive.  Also, as an another example, with $\{\Lambda = -1/30, k_w = 2,  n = 4,\alpha =  M = 1\}$, respecting the positive definiteness of  temperature, the solutions are stable at  $0.645\lesssim r_+\lesssim 0.68$ and $r_+>2.12$, but unstable in $0.9\lesssim r_+\lesssim 1.95$ range.

For this class of solutions, to have Lifshitz rescaling invariant horizon metric, similar to that of  $H^2\times R$ horizon geometry solutions for Einstein field equations with negative $\Lambda$ presented in \cite{Cadeau}, one can set  $a=-\frac{3}{2\Lambda}r_+^{-n}$ or equivalently $\alpha=-\frac{4\Lambda}{3}r_+^n$. Then,  at $M\rightarrow \infty$ limit, when the Ho\v{r}ava-Lifshitz theory tends to general relativity, the thermodynamic quantities are obtained as 
	\begin{eqnarray}\label{ew}
		\begin{split}
			m={\frac {3\sqrt{b}{c}^{3} }{2G\pi \mid \Lambda\mid}}{r_+}^{n}\Omega,\quad
			S={\frac { {c}^{3}}{4G}}A_H,\quad
			T={\frac {8  }{ \pi}}{r_+}^{\frac{n}{2}}
			,
		\end{split}
	\end{eqnarray}
	where $A_H=\frac{3\Omega}{2\mid \Lambda\mid}\sqrt{br_+^{n}}$ and we have eliminated $N_0$ by rescaling of time. The entropy is explicitly in the form of Bekenestein-Hawking entropy form obtained in the solutions of general relativity. 

\subsubsection{A non-Einstein Case:  general $\lambda$ and $n=2$ }
Noting the original form of solutions obtained for general $\lambda$ and $n=2$, given by \eqref{n2iii} and \eqref{n2iii2}, here we consider a different $F(r)$ from that is given by \eqref{Fiii}, as
\begin{eqnarray}\label{}
	\begin{split}
		F(r)=-\frac{\alpha}{6}+C_2 r^2-f(r)	,
	\end{split}
\end{eqnarray}
which leads to
\begin{eqnarray}\label{}
	\begin{split}
		I_E=\frac{\kappa^2}{k_w^4}\beta \Omega\int dtdr N \sqrt {b} \bigg[ &{\frac {
				\alpha   }{144 {M}^{2}{r}^{5}}}\big( -6k_w^2{r}^{3}M{F'}+12 {r}^{2}k_w^{2}MF+6 {M}^{2}{r}^{4}-4 {r}^{2} k_w^2\alpha M+k_w^{4}{\alpha}^{2} \big)\\
		&+
		\frac {1}{\alpha  \left( 4\lambda-1 \right) }
		\bigg(\frac{3}{4r} 
		\left( \lambda-1 \right)  \left( {r}^{2}{{F'}}^{2}+4 
		F ^{2} \right)-\frac{3}{2}  ( 1+
		2 \lambda ) F  {F'} \\
		&+\frac{r}{2}  \left( 3 {C_2}+\Lambda \right)  \big( 3 r{F'
		}-  \left( 3 {C_2}+\Lambda \right) {r}^{2}+6 F \big)  \bigg) \bigg] +B
		,
	\end{split}
\end{eqnarray}
whose equations of motion
gives
\begin{eqnarray}\label{ff}
	F(r)=C_1 r^{-2},\quad N=N_0.
\end{eqnarray}
The variation of the Euclidean action requires the variation of the boundary term   in the following form
\begin{eqnarray}\label{}
	\begin{split}
		\delta B
		=&\frac{-3\sqrt{b} \kappa^2\beta \Omega N}{2 (4\lambda -1)\alpha M k_w^4}\bigg[  \bigg(M r^3 (\lambda -1) F' -MF r^2 (1+2\lambda) +r^4 (3 C_2+\Lambda) M-\frac{1}{36}k_w^2 \alpha ^2(4\lambda -1)\bigg)r^{-2}\delta F  \bigg]^{\infty} _{r_+},
	\end{split}
\end{eqnarray}
where, using  \eqref{ff}, we get
\begin{equation}\label{dbinf2}
	\delta B_{\infty}=-\frac{3}{2}\frac{\sqrt{b}\kappa^2 \beta  N_0\Omega(\Lambda +3 C_2)  \delta C_1}{k_w^4\alpha (4\lambda -1)},
\end{equation}
\begin{eqnarray}\label{}
	\begin{split}
		\delta B_{r_+}= &-{\frac {6\sqrt {b}\Omega   \pi \kappa^2}{r_+^{2} \left( 4\lambda-1 \right)  k_w^4\alpha M}}\big( M \left( -4 {C_2}  \left( 
		\lambda+1 \right) +\Lambda \right) r_+^{4}+\frac{\alpha}{36}
		\left( 4\lambda-1 \right)  \left( 6 r_+^{2}M-{{
				k_w}}^{2}\alpha \right)  \big)
		\delta r_{+}.
	\end{split}
\end{eqnarray}
Also, the temperature based on Euclidean regularity is given  by
\begin{eqnarray}
	\begin{split}
		T=\frac{2 N_0 r _+}{24\pi}  (6 C_2 - r_+).
	\end{split}
\end{eqnarray}
Satisfaction of the first law of thermodynamics by these thermodynamic quantities 
restricts the $C_2$ constant to have one of the following forms
\begin{eqnarray}\label{c22}
	\begin{split}
		C_2^{{{\rm ex}}}= {\frac {\alpha}{12r_+^{2}}},\quad C_2= {\frac {\alpha  \left( 6 r_+^{2}M-\alpha {{
						k_w}}^{2} \right) }{36r_+^{4}M}}
		,
	\end{split}
\end{eqnarray}
where the first expression is identical to the  extremal case. Using $f(r_+)=0$, these two $C_2$ are accompanied by the following $C_1$ constant  
\begin{eqnarray}\label{c}
	\begin{split}
		C_1^{{{\rm ex}}}=-\frac{\alpha r_+^{2}}{12},\quad C_1=- {\frac {{\alpha}^{2} k_w^2}{36M}}.	
	\end{split}
\end{eqnarray}
It is worth mentioning that, in this case the extremal radius of horizon is given in terms of Ho\v{r}ava-Lifshitz parameters as follows
\begin{eqnarray}\label{re}
	\begin{split}
		r_+^{{{\rm ex}}}=\frac{\sqrt{3M\alpha}k_w}{3M}.
	\end{split}
\end{eqnarray}
Now, using the second expressions in \eqref{c22} and \eqref{c}, and performing the integrals we obtain
\begin{eqnarray}\label{}
	\begin{split}
		m=-&{\frac {\Omega {N_0} \sqrt {b}  {c}^{3}}{1152 \pi G\Lambda {
					M}^{2}r_+^{4}}}
		\big( 24 M\alpha r_+^{4} \left( 2 \Lambda k_w^2-3 M \right) \ln 
		\left( r_+ \right) -72 r_+^{6}
		\Lambda {M}^{2}+{\alpha}^{3} k_w^4-18 {\alpha}^{2}{{k_w}}
		^{2}Mr_+^{2} \big)+{ m_0},
	\end{split}
\end{eqnarray}
\begin{eqnarray}\label{}
	\begin{split}
		S={\frac {\sqrt {b}\Omega {c}^{3}  }{8 G\Lambda \alpha M}}
		\left( 4 r_+^
		{3}M\Lambda+6 r_+M\alpha+{\alpha}^{2}{{k_w}}
		^{2}r_+^{-1} \right)+{ S_0}.\quad\quad
	\end{split}
\end{eqnarray}
Also, the heat capacity is given by
\begin{eqnarray}\label{}
	\begin{split}
		C= -&\frac { 
			{c}^{3}\sqrt {b}\Omega}{12\alpha M G\Lambda 
		}\frac { 
			\left( 3 r_+^{2}M-\alpha k_w^2
			\right)}{r_+ 
			\left(r_+^{2}M-\alpha k_w^2 \right) }
		\left( 12 r_+^{4}M\Lambda+6 r_+^{2}M\alpha- k_w^2{\alpha}^{2} \right) 
		,
	\end{split}
\end{eqnarray}
which vanishes when the $r_+$ equals to the following radiuses
\begin{eqnarray}\label{}
	\begin{split}
		&	r_1^2={\frac {\alpha k_w^2}{3M}},\\
		& r_{2,3}^2=-{\frac {  \alpha}{12M\Lambda}}\left( 3 M\mp\sqrt {12 M\Lambda k_w^2+9 {M
			}^{2}} \right),
	\end{split}
\end{eqnarray}
and diverges when
\begin{eqnarray}\label{}
	\begin{split}
		r_+=	\sqrt{3}r_1\equiv r_4.
	\end{split}
\end{eqnarray}
Behavior of heat capacity of this family of solutions  is similar to that of the  Bianchi type $II$ solutions, given by \eqref{Cn2II}.
Temperature is positive definite for $r_+\geqslant r_1$ and  $r_1$ is actually  the extremal radius of horizon introduced by \eqref{re}.   
Depending on the values of  parameters, we can have different behaviors:

(i) If a set of parameters is chosen that makes  ${12 \Lambda M k_w^2+9 {M}^{2}}<0$, there are no real $r_2$ and $r_3$ and the solutions at $r_+ \geqslant r_1$ are stable until $r_+=r_4$ and then become unstable.\footnote{For example with $\{\Lambda =  \frac{1}{40}, M =  \frac{1}{3}, k_w =  4, \alpha = 1\}$ }

(ii) If a set of parameters is chosen that holds  ${12 \Lambda M k_w^2+9 {M}^{2}}>0$, giving $r_2<r_1<r_4<r_3$,   the solutions are unstable {at} $r_1<r_+<r_4$ {region}, and then, after the divergent point $r_4$,  showing stable behavior  at $r_4<r_+<r_3$ range, becomes unstable at $r_+>r_3$ region.\footnote{For example with $\{\Lambda = -0.05, M = 0.5, k_w = 0.6,\alpha = 1\}$.}

(iii) If a set of parameters is chosen that yields  ${12 \Lambda M k_w^2+9 {M}^{2}}>0$, giving
the real radii in the order of  $r_2<r_1<r_3<r_4$, the solutions with $T>0$ are unstable when $r_1<r_+<r_3$,  stable when $r_3<r_+<r_4$ and after divergent point $r_4$ become unstable.\footnote{For example with $\{\Lambda = - \frac{1}{20}, M =  \frac{1}{3},  k_w = 2,\alpha = 1\}$. }

\subsubsection{Non-Einstein Case:  general $\lambda$}

When $\lambda$ and $n$ constants are arbitrary, the solutions for Bianchi type Bianchi type $III$ spacetime are given by \eqref{N2} and \eqref{f2}. There is a resemblance between these solutions and  those of Bianchi type $II$ spacetime, given by \eqref{N2II} and \eqref{f2II}, whose thermodynamic behavior has been studied in section \ref{generallII}. Similarly, we would like to study the thermodynamic behavior of this family of solutions in two cases.

First, we consider the case that $n$ and $\lambda$ are essentially arbitrary and independent, where concerning the asymptotic behavior for $C_1$ and $C_2$ dependent terms in  $f(r)$ function  \eqref{f2}, similar to the solutions of \cite{Cai2009a},  we keep only the $C_1$-dependent terms which has negative power of $r$ for all values of $n$ and $\lambda$. Using the Euclidean action  \eqref{actiontIII}, written in terms of the  $F(r)$ function defined by \eqref{Fiii}, where the  variation of boundary term is given by \eqref{boundariiii},  the solution for the equation of motion of \eqref{actiontIII}
gives the $F(r)$ by 
\begin{eqnarray}\label{Fl1}
	F(r)=C_1 r^{s},\quad N=N_0, \quad s=-\frac{3}{4}(n-2)-\sqrt{\mu},
\end{eqnarray}
where $\mu$ is given by \eqref{mu}.
It can be checked  that in order to have non-vanishing and definite $\delta B_{\infty}$, the constraint of type \eqref{con1} and \eqref{con2} is again required for this Bianchi type solutions.
Applying this condition we get
\begin{eqnarray}
	\delta B_{\infty}={\frac {3\sqrt{b}  \beta {\kappa}^{2}{{n}}^{3}{
				N_0}\left( 3n+2 \right) }{ 32\alpha\left( 5 {n}-2 \right)  k_w^4}}{ C_1} \delta { C_1} \Omega,
\end{eqnarray}
\begin{eqnarray}
	\begin{split}
		\delta B_{+}=&-{\frac {\sqrt {b}\Omega n \left( 3 n+2 \right) \pi
				{\kappa}^{2}}{12M\alpha  k_w^4 \left( 5 n-2
				\right) }}\left( -r_+^{-2 n} k_w^2{\alpha}^{2}+12 \Lambda M+6 r_+
		^{-n}\alpha M \right)
		r_+^{\frac{3}{2} {n}-1}\delta r_+.
	\end{split}
\end{eqnarray}
Also, temperature of this black hole solution is given by
\begin{eqnarray}
	\begin{split}
		T=&{\frac {{N_0}r_+  }{144 M{n}^{2}\pi}} \big(  \left(2- 5 n \right) {\alpha}^
		{2} k_w^2r_+^{-2 n}+6 \alpha M \left( n-2
		\right) r_+^{-n}-12 \Lambda M \left( 3 n+2
		\right)  \big).
	\end{split}
\end{eqnarray}
These thermodynamic quantities satisfy the first law of thermodynamics. Noting that, the $C_1$ constant is given in terms of radius of horizon by
\begin{eqnarray}
	C_1= {\frac { 1}{9M{n}^{2}}}\left( r_+^{-2 n} k_w^2{
		\alpha}^{2}-12 \Lambda M-6 r_+^{-n}\alpha M
	\right) r_+^{\frac{3}{4} n+ \frac{1}{2}},\quad\quad
\end{eqnarray}
the mass and entropy are obtained as follows 
\begin{eqnarray}
	\begin{split}
		m=-{\frac {r_+^{ \frac{3}{2} n+1} {c}^{3} \left( 5 n-2
				\right) {N_0} \Omega \sqrt {b}}{1152 \alpha n{M}^{2} \left( 3
				n+2 \right) \pi G\Lambda}} \bigg(& -{\alpha}^{2}{{k_w
		}}^{2}r_+^{-2 n} \big( 24 \Lambda M+\alpha 
		\left( -r_+^{-2 n}\alpha k_w^2+12 Mr_+^{-n} \right)  \big)\\
		& +36 {M}^{2} \left( 4 \Lambda
		r_+^{-n}\alpha+r_+^{-2 n}{
			\alpha}^{2}+4 {\Lambda}^{2} \right)  \bigg)
		+m_0,
	\end{split}
\end{eqnarray}
\begin{eqnarray}
	\begin{split}
		S= {\frac {\sqrt {b} {c}^{3} \left( 5 n-2 \right)  \Omega}{8\alpha \left( 3 n+2 \right)M G\Lambda}}\big( &4 \Lambda Mr_+^{\frac{3}{2} n}+{r
			_+}^{-\frac{n}{2} } k_w^2{\alpha}^{2}+6 r_+^{\frac{n}{2}}\alpha M \big)
		+S_0
		.
	\end{split}
\end{eqnarray}
Also, the heat capacity for this class of solutions is
\begin{eqnarray}
	\begin{split}
		C=-&{\frac {9 {n}^{3} {c}^{3}\sqrt {b} \left( 5 n-2 \right) \Omega}{
				16\left( 3n +2 \right) G\Lambda \alpha}}
		{  C_1}  r_+^{\frac{3}{4}n- \frac{1}{2}} \big(  \left( 5 n-2
		\right) {\alpha}^{2}k_w^2r_+^{-2 n}+6 
		\left( -\alpha  \left( n-2 \right) r_+^{-n}+2 
		\left( 3 n+2 \right) \Lambda \right) M \big)\\
		&\times\big(k_w^2{\alpha}^{2}
		\left( 5 n-2 \right)  \left( 1-2 n \right) r_+^{
			-2 n}+6 M (2  \left( 3 n+2 \right) \Lambda +\alpha  \left( n-1 \right)  \left( n-2 \right) {r_
			+}^{-n} ) 
		\big)^{-1}
		,
	\end{split}
\end{eqnarray}
which vanishes when $r_+$ equals to either of following radii 
\begin{eqnarray}\label{}
	\begin{split}
		& r_{1,2}^{2n}=-{\frac {\alpha\left(3 M\mp\sqrt {12 \Lambda M k_w^2+9 {M}^{2}}\right)}{12
				\Lambda Mb^{2}}},
		\\
		&	r_{3,4}^{2n}=\frac {1
		}{ k_w^2 \left( 5 n-2 \right) \alpha}\bigg(3 M \left( n-2 \right) \mp\big(9 {M}^{2} \left( n-2 \right) 
		^{2}-12 \Lambda M k_w^2 \left( 15 {n}^{2}+4 n-4 \right) \big)^{\frac{1}{2}}\bigg),
	\end{split}
\end{eqnarray}
and diverges when $r_+$ equals to
\begin{eqnarray}\label{}
	\begin{split}
		r_{5,6}^{-2n}={\frac {1}{ k_w^2  \alpha}}\left( 10 {n}^{2}-9
		n+2 \right)^{-1}\bigg(&3 M \left( {n}^{2}-3 n+2 \right) \\
		&\mp \big(12 \Lambda M k_w^2 \left( 30 {n}
		^{3}-7 {n}^{2}-12 n+4 \right)+9 {M}^{2} \left( {
			n}^{2}-3 n+2 \right) ^{2} \big)^{\frac{1}{2}}\bigg)
		.
	\end{split}
\end{eqnarray}
Similar to what we had in \eqref{r0}, the $r_3$ is not real with $\Lambda<0$. 
In addition, positive definite temperature requires $r_+\geqslant r_4$, where  $r_4$ is the extremal radius of horizon.

To explore the  thermodynamic behavior of heat capacity we choose some values for the appeared parameters in the solutions. 
As an example, setting $\{n=4,\Lambda=-0.02,k_w=M=\alpha=1\}$ that keeps only the $r_1<r_4<r_2$ real,  in the $r_+\geqslant r_4$ region the heat capacity is negative until  $r_+=r_2$ and then becomes positive without any divergence.
Also, as an another example,  if one sets $\{n=4,M=2,\Lambda=-0.01,k_w=0.1,\alpha=1\}$, the order of real radii is $r_1<r_4<r_6<r_5<r_2$.  Here, the solutions are unstable in $r_4<r_+<r_5$ region, then become stable between two divergent points $r_5$ and $r_6$, and then  there is an unstable phase as $r_5<r_+<r_2$, which is followed by a stable phase in $r_+>r_2$ region.

The second group of solutions for \eqref{f2} is indicated by the $f(r)$ function given by \eqref{fIII}, where there is a relation of type \eqref{lanadII}  between $n$ and $\lambda$. 
To study the thermodynamic behavior of this kind of solutions, similar to what  have been done in section \ref{generallII}, we rewrite the action in terms of the new function $F(r)$, defined  by 
\begin{eqnarray}\label{}
	\begin{split}
		F(r)=C_2-\frac{4}{3}\frac{r^2 \Lambda}{n^2}-\frac{2}{3}\frac{\alpha r^{2-n}}{n^2 }+\frac{ \alpha^2 r^{2-2 n} k_w^2}{9n^2  M}-f(r)	,\quad\quad
	\end{split}
\end{eqnarray}
which results in
\begin{eqnarray}\label{}
	\begin{split}
		I_E&=-\beta\Omega\int dr {\frac {{r}^{\frac{3}{2}n}\kappa^2\sqrt {b}\Omega N \left( r \right) 
				\beta}{48 k_w^4\alpha {M}^{2} \left( n+2 \right) } }\bigg[\frac{2}{9} {r}^{-4 n}
		\left( n+2 \right)  \big(  k_w^4{\alpha}^{2}+M
		\left( -6{r}^{n} k_w^2\alpha+9M{r}^{2 n} \right) 
		\big) {\alpha}^{2} \\
		&+{M}
		^{2}{n}^{3} \big( 27 { { \left( 2-n\right)  \left( F-{C_2} \right) ^{2}}{{r}^{-4}}}-36 { {{F'} \left( F-{C_2} \right) }{{r}^{-3}}}+\frac{9}{4} { {{{F'}}^{2} \left( n-2
				\right) }{ \left( n-1 \right) {r}^{-2}}} \big) \bigg]+B
		,
	\end{split}
\end{eqnarray}
whose equation of motion
gives
\begin{eqnarray}\label{3F}
	F(r)=C_1 r^{-\frac{3n}{2}+3},\quad N=N_0.
\end{eqnarray}
The temperature of the black hole based on Euclidean regularity is given  by
\begin{eqnarray}
	\begin{split}
		T=&\frac {N_0 }{72{n}^{2}rM\pi}\big( -k_w^{2}\alpha^2 \left( n+2 \right) r_+^{2-2n}+3M \left(  \left( 9{n}^{2}{ C_2}-2\alpha r_+^{2-n} \right)  \left( n
		2\right) -4r_+^{2}\Lambda \left( 3n-2 \right)  \right) 
		\big).
	\end{split}
\end{eqnarray}
Also, from the variation of the Euclidean action, we find that the variation of the boundary term is given by 
\begin{eqnarray}\label{}
	\begin{split}
		\delta B
		={\frac {3 \sqrt {b}\beta {n}^{
					3}\kappa^2\Omega}{32\alpha  k_w^4 \left( n+2
				\right)  \left( 1-n \right) }}&\bigg[{r}^{\frac{3}{2}(n-2)}\big( 8 \left( n-1\right)  \left( F-{ C_2} \right) -
		r{ F'} \left( n-2 \right)  \big)N\delta F\bigg]^{\infty} _{r_+},
	\end{split}
\end{eqnarray}
which, using \eqref{3F}, leads to
\begin{equation}\label{dbinf}
	\delta B_{\infty}=-\frac{ 3}{4} \frac{\sqrt{b}N_0\beta   \kappa^2 n^3    \Omega}{\alpha k_w^4 (n+2)}C_2\delta C_1,
\end{equation}
\begin{eqnarray}\label{}
	\begin{split}
		\delta B_{r_+}= {\frac {3 \pi  \kappa^2 n\sqrt {b}N_0\Omega}{16\alpha M k_w^4 \left( n+2 \right)  \left( n-1 \right) }}  \bigg(&
		\frac{1}{9} 
		\left( n+2 \right)  \left(2-3 n\right) \left(12 {r}^{2 n}
		\Lambda M-  k_w^2\alpha^2+5\alpha {r}^{n}M \right){r}_{+}^{-\frac{n}{2}-1}\\
		&+\frac{3}{4} M{n}^
		{2}{C_2}  \left( n-2 \right) ^{2}r_+^{\frac{3}{2}(n-2)} \bigg)\delta r_{+}.
	\end{split}
\end{eqnarray}
Satisfactions of the first law of thermodynamics by these thermodynamic quantities 
demands the $C_2$ constant to have one of the following forms
\begin{eqnarray}\label{c2}
	\begin{split}
		C_2=\frac {  r_+^{2}  }{ 27{n}^{2}M	\left(n -2 \right) }\big( & k_w^2\alpha^2 \left( 4-n \right)  r_+^{-2 n}+12M \left(   \left( n-2\right)  \alpha r_+^{-n}+\Lambda \left( 3 n-4 \right)  \right)\\& \pm 2 \left( k_w^2 \left( n-1\right) \alpha^2 r_+^{-2 n}-3 M \left(  \left( n-2 \right) \alpha  r_+^{-n}-4\Lambda \right) \right) \big).
	\end{split}
\end{eqnarray}
Similar to what we had in \eqref{c2II}, the positive sign is not an interesting case since substituting it into $f(r_+)=0$  leads to $C_1r_+^{\frac{3}{2}(n-2)}=0$.
But, with the negative sign,  the expression for $C_2$ constant is identical to the expression  given by the  condition of degenerate horizon     $f(r)=f'(r)=0$ \cite{Myung2010}, where the $C_1$ constant is given by
\begin{eqnarray}\label{extermc1}
	\begin{split}
		C_1^{{\rm ex}}=&\frac {r_+^{ \frac{3}{2}n-1}}{27{n}^{2}M \left(n -2\right) }\big(4\alpha^2 k_w^2 \left( n-1 \right) r_+^{-2n}-12M	\left(  \left( 2-n \right)\alpha r_+^{-n}-4\Lambda \right) \big).
	\end{split}
\end{eqnarray}
Then, we have $T^{{\rm ex}}=0$
and
\begin{eqnarray}\label{Sext}
	\begin{split}
		S^{{\rm ex}}=&{\frac { {c}^{3}\sqrt {b}\Omega }{ 8\left( 3 n-2 \right) G\Lambda M\alpha}
		} \big( 4M \Lambda
		\left( 3 n-2 \right){r_
			+}^{ \frac{3}{2} n} +{\alpha}^{2} k_w^2 \left( n+2 \right) 
		\left( n-1 \right) r_+^{-\frac{n}{2}}+3M  \alpha
		\left( {n}^{2}+4 n-4 \right) r_+^{\frac{n}{2}}  \big).
	\end{split}
\end{eqnarray}
The near horizon geometry
of the above solutions can be found by using the following change of the variables
\begin{eqnarray}\label{}
	\begin{split}
		r\rightarrow \left(r_+^{2-2n}+\frac{\epsilon}{r}\right)^{\frac{1}{2-2n}},\quad t\rightarrow \frac{t}{\epsilon},
	\end{split}
\end{eqnarray}
where sending $\epsilon\rightarrow 0$ and  scaling of time  gives the near horizon metric 
as a product space of $AdS_2 \times H^2\times R$
with different radii
\begin{eqnarray}\label{}
	\begin{split}
		ds^2=\frac{1}{W} &\frac{-dt^2+dr^2}{r^2}+r_+^n\left(a  \left((dx^1)^2+{\rm e}^{2 x^1} (dx^3)^2 \right)+b  (dx^2)^2
		\right),
	\end{split}
\end{eqnarray}
where
\begin{eqnarray}\label{}
	\begin{split}
		W=&{\frac {  {c}^{3}\sqrt {b}\Omega {r_
					+}^{ \frac{3}{2} n}}{ 8\left( 3 n-2 \right) G\Lambda M\alpha}
		}\big( {\alpha}^{2} k_w^2 \left( n+2 \right) 
		\left( n -1\right) r_+^{-2 n}+M \left( 3 \alpha
		\left( {n}^{2}+4 n-4 \right) r_+^{-n}+4 \Lambda
		\left( 3 n-2 \right)  \right)  \big).
	\end{split}
\end{eqnarray}

\section{Conclusion}\label{conclussion}
We have found black hole solutions to $z=4$ Ho\v{r}ava-Lifshitz gravity in $(4+1)$ dimensions, assuming that the horizons possess Bianchi types $II$ and $III$ symmetries. These solutions can be regarded as topological black hole solutions whose negatively curved three-dimensional horizons are modeled on two types of  Thurston's closed $3$-geometries, namely the  Nil geometry and $H^2 \times R$, which are twisted product and product of constant curvature type, respectively. The considered negatively curved geometries do not admit the constant curvature type metric on the horizon, i.e. the Einstein metric $R_{\alpha\beta}=kg_{\alpha\beta}$.
The solutions have been found for  $\beta=-\frac{1}{3}$ in two cases of $\lambda=1$ and general $\lambda$. 
The thermodynamic properties of the solutions have been investigated using the canonical Hamiltonian method.
Interestingly,  except for the differences in the coefficients, the solutions for two Bianchi types $II$ and $III$ have similar forms of metric component functions $f(r)$ and thermodynamic behaviors.

Generally, the solutions and their thermodynamic quantities are given in terms of some constants, including negative cosmological constant $\Lambda$, Ho\v{r}ava-Lifshitz constants $ k_w$, and $ M$, 
	the horizon Ricci scalar dependent parameter $\alpha$ that appears similar to  the $k$ parameter of the topological black hole solutions with constant curvature horizons \cite{L2009,Cai22009,Caia2009,Cai2009},  the constant $n$ of metric that is used to provide distinct classes of solutions, and the two integrating constant $C_1$ and $C_2$ which are not independent and can be given in terms of radius of horizon using the first law of thermodynamics.  Also, the  $a$ and $b$ constants in the metric, which can  provide additional scaling, besides being interpreted in terms of $\alpha$, have been used to establish generalized Lifshitz scaling invariance on the horizon and asymptotic region.

For  $\lambda=1$, one of the interesting outputs of the considered horizon geometries for $z=4$ Ho\v{r}ava-Lifshitz gravity in $(1+4)$ dimensions was existence of a logarithmic branch in the solutions for metric.
	Even though, similar to the other $(1+4)$ dimensional $z=4$ black hole solutions, their entropy does not contain logarithmic correction that appears in $z=3$ black hole solutions in $(1+3)$ dimensions. Also, it has been shown that when the Ho\v{r}ava-Lifshitz terms are neglected at $M\rightarrow \infty$ limit, the $\lambda=1$ solutions can behave similar to the solutions obtained for vacuum Einstein equations with negative cosmological constant  \cite{Cadeau,PhysRevD.91.084054,Nil1}.

For general $\lambda$, we first considered the special case of $n=2$. In addition, allowing  $n$ to primarily   have  arbitrary value, we considered two cases using the asymptotic behavior of the solutions and imposing relations between $n$ and $\lambda$. Then, the appearing $n$ constant in these classes of solutions refers actually to the general value of $\lambda$.
It turned out that for the solutions {that possess} two integrating constants  $C_1$ and $C_2$, the energy in Hamiltonian formalism requires more information to be integrated, and the two integrating constants need to be subject to a boundary condition imposed as a functional relation between  $C_1$ and $C_2$. 
In fact, this feature and the necessity of imposing the boundary condition were first observed in Einstein-Scalar theory \cite{Hertog}, where in the Hamiltonian formalism of mass the functional relation between dilaton charge and the dilaton asymptotic value is required, which can be fixed uniquely if the asymptotic AdS symmetry is of interest \cite{Henneaux2007,Anabal2015,Anabalon2016},  leading to satisfaction of first law without including the variation of the non-physical charge of Dilaton \cite{ASTEFANESEI201847,Naderi2019}. Here, having only the mass as the physical characteristic of the black hole solutions, we employed the first law of thermodynamics to determine the suitable functional relation between $C_1$ and $C_2$, which enabled us to calculate the well-defined mass in terms of the radius of the horizon.

A generic feature of  the obtained solutions  is that the entropy for all classes of solutions with both  horizon geometries of Nil and $H^2\times R$, besides  containing a term proportional to the area of horizon $A_H$, receive  two negative corrections of type $A_H^{\frac{1}{3}}$ proportional to $\Lambda^{-1}$, and  $A_H^{-\frac{1}{3}}$ proportional to Ho\v{r}ava-Lifshitz constant $k_w$. The latter one, which  shows a divergent at $r_+\rightarrow 0$ limit, is a particular consequence of the considered unusual horizon geometries and does not appear in the entropy of $(1+4)$ dimensional topological black hole solutions of $z=4$ Ho\v{r}ava-Lifshitz gravity with spherical and hyperbolic horizons, obtained in \cite{Cai22009,Liu2013}. However, similar to the constant curvature horizon $(1+4)$ dimensional $z=4$ solutions presented in \cite{Cai22009,Liu2013}, the entropy for our obtained solutions with   horizon geometries of Nil and $ H^2 \times R$ did not receive  logarithm
correction that is common in $(1+3)$ dimensional black hole solutions in Ho\v{r}ava-Lifshitz gravity \cite{Cai2009,Myung2010,Myung2009}.  Furthermore, investigating the behavior of heat capacity, it is found out that with proper choices of parameters, the locally stable or unstable phases can appear for all classes of solutions.

In addition, classes of extremal black hole solutions have been provided for both Bianchi types $II$ and $III$ models.
We have shown that with general $\lambda$ if there is a relation of type \eqref{con1} between $\lambda$ and $n$, the consistent solutions are restricted to be in the extremal cases. The near horizon geometries for these extremal black holes were obtained as $AdS^2\times Nil$ and $AdS^2\times H^2\times R$ for Bianchi types $II$ and $III$ solutions, {respectively}.
These solutions possess finite entropy at
zero temperature,  similar to extreme near horizon Reissner-Nordstrom black hole solution. 
This is also similar to the behavior of Bianchi type $II$ and $III$ charged black hole solutions in extremal near horizon limit that we have studied in \cite{Naderi2019} in the context of string theory.

The $(4+1)$ dimensional black hole solutions for $z=4$ Ho\v{r}ava-Lifshitz gravity with flat, hyperbolic and spherical {horizons} have been already studied in \cite{Cai2009,Chench2009,Liu2013}. There is a correspondence between these geometries and Bianchi types $I$, $V$ (isotropic expansion), and $IX$. 
It would be {also} desirable to investigate black hole solutions for Ho\v{r}ava-Lifshitz gravity with horizons modeled on the other Thurston type geometries of $\widetilde{{SL_2R}}$ and solve geometry, which correspond to the homogeneous spaces with Bianchi types  $VIII$ and $VI_{-1}$ symmetries. A difficulty in finding solutions with these symmetries is that the equations of motions contain higher derivative terms that can not be eliminated by suitable choices of constants, however, further work is under progress in this sense. Also, in view of applications of black holes with Thurston horizon geometries in AdS/CFT context, 
	where the symmetry requirements on spatial directions are slightly relaxed considering homogeneity instead of usual translational symmetries
	\cite{Iizuka2012,Nil1}, it would be interesting to
	further analyze the $(1+4)$ dimensional  black holes we obtained for $z=4$ Ho\v{r}ava-Lifshitz gravity.

\appendix
\section{Appendix}\label{app1}
In this appendix,  we present  Ricci scalars and the components of $R_{ij}$, $K_{ij}$, and $L_{ij}$ tensors   for both  Bianchi types $II$ and $III$ models, where the $i$ and $j$ indices run over the radial coordinate $r$ and the Bianchi space part indices $\{1,2,3\}$.

\subsection{Bianchi type $II$}
For this Bianchi type, with the considered metric ansatz \eqref{metricII},  
the non-zero component of Ricci tensor are given by
\begin{eqnarray}\label{rnon}
	\begin{aligned}
		R_{rr}&=-\frac{1 }{4r^2 f} \left(f'  r(2   m +  n)+2m f( m -2) + n f(n  -2)\right),\\ 
		R_{11}&=-\frac{1 }{4b r^{m+2} }\left(f'r^{2m+1}  b^2 m +2 r^{2m} f b^2 m(m-1)+n  f b^2 r^{2m} m+2 a r^{n+2}\right),\\
		R_{22}&=-\frac{a r^n}{4b^2  r^{2m+2}} \left(  f' r^{2m+1} b^2 n +2 n f b^2 r^{2m}( m-1)+r^{2m} f b^2 n^2 -2 a r^{n+2} \right), \\
		R_{23}&=\frac{ a r^n x^1}{4b^2 r^{2m+2} } (f' r^{2m+1} b^2 n +2 n f b^2 r^{2m} (m-1)+r^{2m} f b^2 n^2 -2 a r^{n+2} ) ,\\
		R_{33}&= \frac{1}{ 4  r^{ 2}}(-   (a n (x^1)^2 r^n+b m r^m)rf' -  f (2m+1-2)\left(  m b  r^{ m} +a n  (x^1)^2   r^{ n}\right) \\
		&~~~~~~~~~~~~~~~~~~~~~~~~~~~~~  -2 a r^{n+2}( b^{-1}r^{-m}-  ab^{-2}r^{n-2m} (x^1)^2)).
	\end{aligned}
\end{eqnarray}
Also, the Ricci scalar is
\begin{eqnarray}
	R=\frac{1}{2b^2  r^{2m+2}}\left(- b^2 (2m+n) r^{2m+1} f' -  (3m^2+({2}n-{4}) m+n^2-{2}n) b^2 f r^{2m } -a r^{n+2}\right).
\end{eqnarray}
The extrinsic curvature tensor $K_{ij}$, defined by \eqref{Kij}, does not have non-zero components with the metric \eqref{metricII}. Also, for $L_{ij}$ defined by \eqref{Lij}, we have the following components, considering $\beta=-\frac{1}{3}$
\begin{eqnarray}\label{}
	\begin{aligned}
		L_{rr} =\frac{1}{48f b^4  r^{4m+4} } \big(	b^4 r^{4m+2}  (m-n)^2(&-4 f f''+  f'^2-4 r^{-1} f   (m-2+\frac{n}{2}) f')
		-  (n-2) f ^2 (4m-n-6)   r^{-2}\\
		& +20 a b^2 r^{n+2m+2} f (m-n)^2 -16 r^{2n+4} a^2  \big),
	\end{aligned}
\end{eqnarray}
\begin{eqnarray}\label{}
	\begin{aligned}
		L_{11}= \frac{1}{48}&\frac{1}{b^3 r^{3m+4}}\big(r^{4m+2} b^4 (m-n)\big(-4 r f  f'''-8   (\frac{1}{4}r  f'+f (m+\frac{3}{4}n-2))  f''
		-3   (m+\frac{1}{3}n-2) f'^2 \\
		&+   20 r^{-2m-1} b^{-2} (-\frac{1}{5}f b^2 (m^2+(\frac{7}{2}n-9) m-\frac{11}{2}n+12) r^{2m} +a r^{n+2} )    f'\\
		&- f  b^{-2} r^{-2} (  (2-n) f b^2  (n-4+m)   ( n + 6-4m) +5 a   r^{-2m+n+2}   (m-2 n+2))\big)+48  a^2 r^{2n+4} \big),
	\end{aligned}
\end{eqnarray}	
\begin{eqnarray}\label{}
	\begin{aligned}
		L_{22}= -\frac{5}{3}\frac{r^n  a}{b^4 r^{4m+4}}&  \big( b^4 (m-n)\big(-\frac{1}{10}r^{4m+3} f f^{\prime \prime \prime} - \frac{1}{4} r^2 (\frac{1}{5} r f'+f (m+\frac{2}{5}n-\frac{8}{5})) r^{4m}  f''\\
		&+\frac{1}{2} b^{-2} r (-\frac{3}{10} b^2 f (-\frac{1}{6} n^2+(\frac{13}{6}m-3) n+m^2-\frac{20}{3} m+8) r^{2m} +a r^{n+2} ) r^{2m} f'\\
		&  -\frac{1}{16}  (m+\frac{3}{5}n-\frac{12}{5}) r^{4m+2} f'^2 -\frac{3}{20} (m-\frac{3}{2}-\frac{1}{4}n) f ^2 (n-2) (m-\frac{8}{3}+\frac{1}{3}n) r^{4m}\\
		& -\frac{1}{ 4} r^{2m+n+2} f a b^{-2}   (m-3 n+4)\big)  +r^{2n+4} a^2  \big),
	\end{aligned}
\end{eqnarray}	
\begin{eqnarray}\label{}
	\begin{aligned}
		L_{23}= \frac{5}{3}\frac{a r^n x^1}{b^4  r^{4m+4}}&   \big(b^4 r^{4m } (m-n)\big(-\frac{1}{10} f   r^{ 3}   f^{\prime \prime \prime} -\frac{1}{20} r^2 ( f' r+ 5 (m+\frac{2}{5}n-\frac{8}{5}) f )   f''\\
		&+\frac{1}{2}(-\frac{3}{10}(-\frac{1}{6}n^2+(\frac{13}{6} m-3) n+m^2-\frac{20}{3}m+8) b^2 f r^{2m} +a   r^{n+2}) r^{-2m+1}  b^{-2}  f'\\
		& -\frac{1}{16}r^2 (m+\frac{3}{5}n-\frac{12}{5})  f'^2 - \frac{3}{20} (m-\frac{3}{2}-\frac{1}{4}n) f ^2 (n-2) (m-\frac{8}{3}+\frac{1}{3}n)  \\
		& -\frac{1}{4}r^{n+2} f a b^{-2}  (m-3 n+4) r^{-2m} \big)+ r^{2n+4} a^2 \big),
	\end{aligned}
\end{eqnarray}		
\begin{eqnarray}\label{}
	\begin{aligned}
		L_{33}=\frac{1}{48}&\frac{1}{b^4 r^{4m+4}}\big( (m-n) b^4 r^{4m } \big(8 (r^{n } (x^1)^2 a-\frac{1}{2}r^m b) r^{ 3}f^{\prime \prime \prime}+ 4   (r (r^n (x^1)^2 a-\frac{1}{2}r^m b)f'  \\
		&+(-2b (m+\frac{3}{4}n-2) r^m+r^n (x^1)^2 a (5m+2n-8)) f ) r^{ 2} f''+  r^{ 2} (-b (3m+n-6) r^m\\
		&+r^n (x^1)^2 a (5m+3n-12)) f'^2-f'  b^{-2} r^{-2m+1} (4(m^2+(\frac{7}{2}n-9) m-\frac{11}{2}n+12) b^3 f r^{3m} \\
		&-12 a (m^2+(\frac{13}{6}n-\frac{20}{3}) m-\frac{1}{6}n^2-3 n+8) r^{2m+n} (x^1)^2 b^2 f -20a b r^{m+n+2} +40a^2 r^{2n+2} (x^1)^2) \\
		& -   (-n-6+4m) b  (n-2) (n-4+m) f^2 r^{ m} + a r^n   (-n-6+4m) (3m-8+ n) (x^1)^2
		(n-2) f ^2 r 
		\\
		&	-20 a b^{-1} r^{n- m+2} f   (m-2 n+2) +20 a^2 b^{-2} r^{2n-2m+2} (x^1)^2 f  (m-3 n+4)\big) \\
		&+48 r^{m+2n+4} a^2 b -80 r^{3n+4} a^3 (x^1)^2\big).
	\end{aligned}
\end{eqnarray}

\subsection{Bianchi type $III$}

For Bianchi type $III$, with the metric ansatz \eqref{metricIII}, the non-zero components of Ricci tensor are
\begin{eqnarray}
	\begin{aligned}
		R_{rr}=- \frac{1 }{ 4f r^2}\left(r (m+2 n)f' +f (m^2+2 n^2-2 m -4 n)\right),\\ 
		R_{11}=-\frac{1}{4r^2} \left(f'a r^{n+1} n  +a n f (m+2 n-2) r^n+4 r^2\right),\\
		R_{22}=-\frac{bm}{4} r^{m-2}\left( f' r+f (m+2 n-2)\right),\\
		R_{33}=-\frac{e ^{2 x^1} }{4r^2}(f' a r^{n+1} n  +a n f (m+2 n-2) r^n+4 r^2).
	\end{aligned}
\end{eqnarray}
The Ricci scalar is given by
\begin{eqnarray}
	R=-\frac{1}{2a r^{n+2} }(a   r^{n+1} (m+2 n) f' +f a (m^2+(2 n-2) m+3 n^2-4 n) r^n+4 r^2).
\end{eqnarray}
The
extrinsic curvature $K_{ij}$, defined by \eqref{Kij},  vanishes with metric \eqref{metricIII}, and the non-zero components of tensor $L_{ij}$, defined by \eqref{Lij}, considering $\beta=-\frac{1}{3}$, are as follows
\begin{eqnarray}\label{}
	\begin{aligned}
		L_{rr} = \frac{1}{48} \frac{1}{f r^{2 n+4} a^2}\big(	 (m-n)^2  r^{2 n} \big(-4 a^2& r^{ 2} f   f''+a^2 r^{ 2}   f'^2-2 a^2 r  f (m-4+2 n)   f'\\
		&	+a^2 f^2 (m-2) (m-4 n+6)    \big) -16 r^4\big),
	\end{aligned}
\end{eqnarray}
\begin{eqnarray}\label{}
	\begin{aligned}
		L_{11}= \frac{1}{48}\frac{1}{a r^{n+4}} \big(&r^{2n }(m-n)\big(4 a^2 r^{ 3} f   f''+6  a^2 r^{ 2} f^{\prime \prime}  (\frac{1}{3}r f'+f (m+\frac{4}{3}n-\frac{8}{3})) \\
		&+a^2 r^{ 2}  (m+3 n-6)  f'^2+ 14  r  a^2 f  f' (\frac{2}{7}n^2+(m-\frac{18}{7}) n-\frac{11}{7} m+\frac{24}{7}) \\
		&-a^2 f ^2 (m-2) (-4 n+6+m)  (n-4+m)  \big)+16 r^4 \big),
	\end{aligned}
\end{eqnarray}	
\begin{eqnarray}\label{}
	\begin{aligned}
		L_{22}= -\frac{r^m b }{6a^2 r^{2n+4} }\big( (m-n) r^{2n} &\big(a^2 r^{ 3} f   f'''+(\frac{1}{2} r f'+f (m+\frac{5}{2} n-4)) r^{ 2} a^2  f''+\frac{3}{8} r^{ 2} a^2  (m+\frac{5}{3} n-4) f'^2\\
		&-\frac{1}{4}   a^2 (m^2+(-13 n+18) m-6 n^2+40 n-48) f  r f' \\
		&-\frac{1}{8}f ^2 a^2 (m-2) (-4 n+6+m)  (3 n-8+m)\big) +2 r^4\big),
	\end{aligned}
\end{eqnarray}	
\begin{eqnarray}\label{}
	\begin{aligned}
		L_{33}= \frac{e ^{2 x^1}}{8a r^{n+4} }&\big(	 (m-n)  a^2   r^{2n} \big(\frac{2}{3}f r^{ 3}  f^{\prime \prime  \prime}+ r^{ 2}    (\frac{1}{3}r f' +f (m+\frac{4}{3} n-\frac{8}{3})) f''\\
		& +\frac{1}{6}  r^{ 2} (m+3 n-6)  f'^2+ \frac{1}{3} r    f  f' ({2}n^2+(7m-{18}) n-{11}m+{24}) \\
		& -\frac{1}{6}f ^2   (m-2) (-4 n+6+m)   (n-4+m)  \big) +\frac{8}{3}r^4\big).
	\end{aligned}
\end{eqnarray}

 
	
\bibliographystyle{h-elsevier}
\bibliography{blackholehoeava1}
\end{document}